\documentclass{iopart}
\usepackage{epsfig}
\usepackage{latexsym}
\begin{document} 
 
\newcommand{\tbox}[1]{\mbox{\tiny #1}} 
\newcommand{\half}{\mbox{\small $\frac{1}{2}$}} 
\newcommand{\mbf}[1]{{\mathbf #1}} 
\newcommand{\mpg}[3][b]{\begin{minipage}[#1]{#2}{#3}\end{minipage}}
 
 
\title{Consolidating boundary methods 
for finding the eigenstates of billiards}
 
\author{Doron Cohen$^{1}$, Natasha Lepore$^{2}$ and Eric J. Heller$^{2,3}$} 
  
\address{ 
\mbox{$^1$ Department of Physics, Ben-Gurion University, 
      Beer-Sheva, Israel.} \\
\mbox{$^2$ Department of Physics, Harvard University, 
      Cambridge, Massachusetts.} \\ 
\mbox{$^3$ Department of Chemistry and Chemical Biology, 
      Harvard University, Cambridge, Massachusetts.} \\ 
}

 
\begin{abstract} 
The plane-wave decomposition method (PWDM), a widely used
means of numerically finding eigenstates of the Helmholtz equation in 
billiard systems is described as a variant of the mathematically 
well-established boundary integral method (BIM). A new unified 
framework encompassing the two methods is discussed. Furthermore, 
a third numerical method, 
which we call the Gauge Freedom Method (GFM) is derived from the BIM 
equations. This opens the way to further improvements in eigenstate
search techniques. 
\end{abstract} 
 
 
\vspace*{-0.5cm}
 
\section{Introduction} 
 
Solving the Helmholtz equation within a domain 
given Dirichlet boundary conditions 
is of great interest to both physicists \cite{sto} and
engineers. Firstly, the Helmholtz equation is the simplest 
example of a \mbox{\em wave equation}. Furthermore, this 
equation may be used to describe acoustics waves, microwave 
systems, and in particular the wavefunction 
of a quantal particle inside nano-scale devices \cite{datta}  
such as quantum-dots, where the motion of the electrons 
can be regarded as a free~motion within a box. For this reason   
it has become a prototype problem in studies of quantum chaos. 

Of particular interest are the wavefunctions $\Psi(x)$ of a
stationary particle in a two-dimensional box (a so-called 
billiard system). These wavefunctions are 
solutions of the homogeneous Helmholtz equation 
\mbox{${\cal H}\Psi(x) = 0$}, where 
the differential operator ${\cal H}$ is defined as 
\begin{eqnarray} \label{e1} 
{\cal H} = -\nabla^2 - k^2. 
\end{eqnarray} 
Note that for the special case $k=0$, the Helmholtz equation 
reduces to Laplace's equation. Given a closed boundary 
we can ask whether this equation has a non-trivial solution 
that satisfies Dirichlet boundary conditions $\Psi(x)=0$. 

Two main numerical strategies have been suggested to date in the 
literature in order to find the eigenstates of the Helmholtz equation 
(for more comprehensive reviews and references, see 
for example \cite{kuttler,backer,barnett}).  
The first strategy can be described as a `Laplacian diagonalization'.  A basis 
is selected such that the functions it contains satisfy the Dirichlet 
boundary conditions. For example, in some cases one can use conformal 
mapping to determine a basis \cite{conformal} (and see also 
\cite{reichl}). The Laplacian operator is then written in this 
basis and diagonalized. Numerically, some truncation 
is required, and the diagonalization only determines all the 
eigenstates up to some maximum wavenumber $k_{\tbox{max}}$.  
Thus, the Laplacian diagonalization strategy is inherently limited, 
and can not be used for the purpose of finding high-lying eigenstates.

The second numerical strategy, which is the object of this paper, can be 
described as a `boundary approach'. This strategy is based on the 
observation that the eigenfunctions are completely determined by their 
behavior at the boundary. 
The boundary methods use basis functions that satisfy the Helmholtz equation 
inside the billiard at fixed $k$. A linear combination of 
the basis functions is then
selected such that the boundary conditions are satisfied. 
Thus, in order to find the eigenstates, one only needs to study the 
small $k$~window that contains the energy range of interest. 
Therefore the method is naturally suitable for the purpose of 
finding high-lying eigenstates. 
For 2D~billiards, the Laplacian diagonalization 
requires 2D~grid calculations. This is a heavy numerical task.  The 
boundary approach, on the other hand reduces the calculations to a 
1D~boundary grid.

In the quantum chaos community, two boundary methods are commonly 
employed. The first one is referred to as the boundary integral method 
(BIM) \cite{berry1}, while the other is what we call here 
the decomposition method (DEM), of which the 
plane-wave decomposition method (PWDM) \cite{heller2}
is a special case. Extensions of the standard PWDM have 
been used in \cite{vergini,VS} and in \cite{barnett}. 

Usually, the BIM and the PWDM are considered to be two
independent self-contained procedures. Several studies
have been done in order to compare their capabilities \cite{li}. 
While the BIM equation is exact, its convergence is very slow 
(power law in the number $b$ of discretization points per half-wavelength). 
On the other hand while the PWDM is mathematically 
limited (e.g. the maximal $b$ is semiclassically determined), 
it is still found to be extremely efficient in practice. 
Hence there is definitely a need to develop hybrid boundary methods.


In the present paper, we adopt a new point of view through 
which we regard the BIM and the DEM as sequences of 
four independent steps. By doing so, we are going to make the observation 
that the DEM and the BIM are strongly related: {\em The two procedures
are based on the diagonalization of literally the same matrix!}
As a bridge between them, we will highlight an intermediate strategy
which we call the gauge-freedom method (GFM). In a follow-up paper,
this framework will lead the way to improved eigenstate 
search techniques combining the strengths of the two boundary
methods \cite{bmi}.
 
Our unified description of the different boundary methods 
can be summarized by the following set of four steps that are 
common to the BIM and the DEM, and as we will show later, to the GFM: 
%
\begin{itemize} 
\item Choice of a set of basis functions $F_j(x;k)$. 
\item Definition of the Fredholm matrix $\mbf{A}_{js}(k)$. 
\item Procedure for construction of the wavefunction $\Psi_r$. 
\item Definition of the quantization measure $S(k)$. 
\end{itemize} 

The first step consists of selecting a set of basis 
functions $F_j(x;k)$ labeled $j=1..N$. 
All boundary methods rely on basis functions that 
satisfy the Helmholtz equation {\em inside} the billiard. 
Thus, a superposition of such basis functions is an 
eigenfunction if it vanishes along the boundary. 
The choices of bases that correspond to the PWDM, 
to the primitive version of the BIM and to the simplest 
variation of the GFM are as follows:  
\begin{eqnarray} \label{e2} 
F_j(x;k) \ \ =& \ \ \cos(\phi_j + k {n}_j \cdot {x}) 
\ \ & \ \ \ \ \mbox{PWDM} 
\\ \label{e3} 
F_j(x;k) \ \ =& \ \ Y_0(k|{x}-{x}_j|) 
\ \ & \ \ \ \ \mbox{Y0-BIM} 
\\ \label{e4} 
F_j(x;k) \ \ =& \ \ J_0(k|{x}-{x}_j|) 
\ \ & \ \ \ \ \mbox{J0-GFM} 
\end{eqnarray}

For the purpose of the numerical treatment we represent 
the boundary by a set of points $x_s$ with $s = 1\cdots M$. 
In practice, we choose a set of $M$ equally spaced points, 
so that the spacing is $\Delta s = L/M$ where $L$ is the 
perimeter of the billiard. Depending on details of
the numerical strategy, 
the number of points along the boundary is either taken to 
be equal to the number of basis functions ($M=N$), 
or it may be larger ($M>N$). The Fredholm matrix is defined as 
\begin{eqnarray} \label{e5} 
\mbf{A}_{js}(k) \ \ \equiv \ \ F_j(x_s;k) 
\end{eqnarray} 
Given $k$, one may perform the singular value decomposition (SVD) 
of the matrix $\mbf{A}$. 
The smallest singular value is the
one which we care about. If it is a minimum at a given $k$, 
then the billiard system is likely to have an eigenvalue at that energy.

In the third step, the left and right
eigenvectors of the smallest singular value 
($\Phi_s$ and ${\mathsf C}_j$, resp.) 
are used to construct a wavefunction $\Psi_r$ through a linear 
transformation. We select a grid of points $X_r$ on which 
the wavefunction $\Psi_r \equiv \Psi(X_r)$ is calculated. 
In the DEM, the left eigenvector ${\mathsf C}$ is used for
the purpose of this construction, and the linear transformation 
which is applied is: 
\begin{eqnarray} \label{e6} 
\Psi_r = \sum_j {\mathsf C}_j \mbf{F}_{jr} 
\end{eqnarray} 
where $\mbf{F}_{jr} \equiv F_j(X_r;k)$. 
Note that ${\mathsf C}$ contains the expansion coefficients 
of $\Psi(x)$ in the chosen basis $F_j(x;k)$. 
For the BIM, the right eigenvector $\Phi$ 
is used in order to build the wavefunction, 
and the linear transformation in this case is 
\begin{eqnarray} \label{e7} 
\Psi_r = \sum_s \mbf{G}_{rs} \Phi_s, 
\end{eqnarray} 
where $\mbf{G}_{rs}$ is the discretized version of 
the Green function. Thus, the vector $\Phi_s$ 
represents a `charge' that is distributed along the boundary.  
 
In the final step, a measure $S(k)$ is defined such that 
$S(k)=0$ if $k$ is an eigenvalue and $S(k)>0$ otherwise. 
In practice, the eigenvalues are determined by searching for the local 
minima of $S(k)$. By construction, the wavefunction which was built 
in step three satisfies the Helmholtz equation inside the boundary. 
Therefore, the most natural choice of $S(k)$ is the tension,
the sum of the square of the wavefunction along the boundary. 
The tension is thus a
measure for the roughness of the constructed $\Psi(x)$ 
along the boundary. This definition of $S(k)$ is 
traditionally used with the PWDM. 
Other possibilities for the measure include 
the smallest singular value, and the Fredholm 
determinant of $\mbf{A}$. These two latter choices 
of $S(k)$ are the ones that are usually associated with the BIM. 
In Sec.~\ref{IV} we discuss the mathematical equivalence 
of the three possible measures, and compare 
their respective numerical effectiveness.

\subsection{Outline} 

In Sec.~\ref{II}, we give a concise presentation 
of the BIM and the related GFM. Our derivation of 
the BIM equation contains some significant 
improvements over previous ones. Most
importantly, it naturally leads to the 
existence the GFM.
Furthermore, we have succeeded in avoiding the use 
of the complicated "regularized" method of images, 
which was the major ingredient in the derivation 
of Ref.~\cite{li}. 
 
Strategies for constructing the wavefunction 
are discussed in Sec.~\ref{III}. An explanation of 
the Green function method is given, 
as well as a critical discussion of the DEM and 
its numerical variants. 
 
Sec.~\ref{IV} explores the practicality of using different 
choices for the quantization measure. In particular, 
it is demonstrated that a tension measure can be defined 
not only for the PWDM, but for the case of the BIM as well.  
An important issue emerges as to whether the 
quantization measures can be used as to determine
the error in the bulk wavefunction.  
We address this issue, and also make a comparison 
between the numerical accuracies of the BIM and of the PWDM. 

Sec.\ref{V} explains how the GFM bridges between 
the BIM and the DEM. It is found that for 
any DEM, an associate GFM exists, whereas the 
inverse statement is not true.

The shape that we have studied numerically is presented in Fig.1. 
We have used the cornerless, generic 'Pond' shape in order to avoid 
the range of problems that arise with more complicated geometries. 
These problems are the subject of a follow-up study \cite{bmi}, 
where we suggest mixed BIM/DEM methods for finding eigenfunctions. 
This is done using the above theoretical framework, 
while regarding the 'Pond' shape as a reference against which to judge 
the effectiveness of our efforts. Another direction of 
research is related to billiards in magnetic fields \cite{klaus}. 
  
For the convenience of the reader, our 
numerical notations are concentrated in Table~1. 
Further information about Fig.~1, Table~1, 
and the numerical analysis is integrated within 
the main text.

\begin{table} 
\begin{tabular}{|lll|} 
\hline 
\ & \ & \\ 
$x_s$ &=& vector of boundary points \hspace*{3cm} \\ 
$x_{{\mathsf s}}$ &=& vector of outer-boundary points \\ 
$X_r$ &=& vector of interior grid points \\ 
$X_0$ &=& randomly selected interior point \\ 
\ & \ & \\ 
$\Psi_r$ &=& wavefunction on the grid points \\
$\Psi_s$ &=& wavefunction on the boundary points \\ 
$\Phi_s$ &=& `charge' along the boundary \\ 
\ & \ & \\ 
$\| \Psi_r \|$ &=& norm of the wavefunction (see Sec.III) \\ 
$\| \Psi_s \|$ &=& tension of the wavefunction (see Sec.III) \\ 
\ & \ &  \\ 
$n(s)$  &=& unit normal at the boundary point $x_s$ \\ 
\  & \ & \\ 
$\mbf{w}_{s}$  &=& $(1/(2k^2)) \ n(s) \cdot  x_s $ \\ 
\  & \ & \\ 
$\mbf{G}_{rs}$ &=& $G(X_r,x_s)$ \\ 
\  & \ & \\ 
$\mbf{A}_{j0}$ &=& $F_j(X_0;k)$ \\ 
\  & \ & \\ 
$\mbf{A}_{js}$ &=& $F_j(x_s;k)$ \\ 
\  & \ & \\ 
$\mbf{D}_{js}$ &=& $\partial F_j(x_s;k)$ \\ 
\ & \ &  \\ 
$\mbf{B}_{ij}$ &=& $\Delta s \ \sum_s \mbf{w}_s \mbf{D}_{is}\mbf{D}_{js} 
                    \ \ = \ \ 
            \Delta s \ (\mbf{D}\mbf{w}\mbf{D}^{\dag})_{ij}$ \\ 
\  & \ & \\ 
$\mbf{T}_{ij}$ &=& $\Delta s \ \sum_s \mbf{A}_{is} \mbf{A}_{js} 
                    \ \ \ \ = \ \ 
            \Delta s \ (\mbf{A} \mbf{A}^{\dag})_{ij}$  \\ 
\  & \ & \\ 
\hline 
\end{tabular} 
\caption{Notations} 
\end{table} 

\section{The BIM and the GFM} 
\label{II}

The gist of the BIM is that, 
from the knowledge of the gradient of the wavefunction 
on the boundary and from Green's theorem, it is possible 
to find the value of the wavefunction everywhere 
inside the billiard. We give a derivation of this method 
in this section. This procedure will lead us naturally 
to the existence of the GFM.  

The free space Green function $G(x,x')$ is defined 
by the equation ${\cal H}G(x,x')=\delta(x-x')$. 
The most general solution can be written as 
\begin{eqnarray} \label{e8} 
G(x,x') \ \ = \ \ - \frac{1}{4} Y_0(k|x-x'|) 
\ \ + \ \ {\cal C}(x,x') 
\end{eqnarray} 
where ${\cal C}(x,x')$ is any solution of 
the homogeneous equation ${\cal H}{\cal C}(x,x')=0$. 
Note that in the electrostatic limit $k\rightarrow0$ 
we have $G(x,x')=-(1/(2\pi))\ln(r) + {\cal C}$, 
where ${\cal C}$ is a constant~\cite{rmrk2}. 
We shall refer to the choice of ${\cal C}(x,x')$ 
as gauge freedom. This term is 
at the core of the GFM.
 
By the definition of the Green function, it follows 
that a solution of the generalized Poisson-Laplace (GPL) 
equation ${\cal H}\Psi(x) = \rho(x)$ is~\cite{rmrk3} 
\begin{eqnarray} \label{e9} 
\Psi(x) \ \ = \ \ \int G(x,x')\rho(x') dx' 
\end{eqnarray} 
We refer to $\rho(x)$ as the `charge density', 
by analogy to its electrostatic equivalent. 

We shall use the notation $\Phi(s)$ in order to refer 
to the (surface) charge density upon the boundary. 
In the latter case, the equation above reduces to 
\begin{eqnarray} \label{e10} 
\Psi(x) \ \ = \ \ \oint G(x,x(s)) \Phi(s) ds 
\end{eqnarray} 
where $s$ parameterizes the boundary. 

\subsection{The BIM} 
 
Let us assume that $k$ is an eigenvalue of the billiard. 
In such a case, there exists a non-vanishing $\Psi(x)$ 
inside the boundary that satisfies $\Psi(x)=0$ on the boundary. 
It can be shown from Green's theorem that the interior 
wavefunction satisfies Eq.(\ref{e10}) with 
\begin{eqnarray} \label{e11} 
\Phi(s) \ = \  \partial_{-} \Psi(x(s)) \ \equiv \ 
\lim_{x\uparrow x(s)} n(s)\cdot\nabla \Psi(x) 
\end{eqnarray} 
where $n(s)$ is the outward pointing normal at point $s$, and 
$\partial_{-}$ is used for the normal derivative evaluated inside
of the billiard walls. 

In electrostatics, it is known that forcing the 
scalar potential to be zero on the boundary induces 
a boundary charge. From Green's theorem,
the induced charge is proportional to the normal
component of the electric field. Here the wavefunction 
acts as the equivalent of the scalar potential. Similarly 
to the electrostatic case, there exists an induced
`boundary charge', that is in this case proportional 
to the normal derivative of the wavefunction.
 
The BIM is based on the fact that if an eigenstate exists, 
then there also exists a charge density $\Phi(s)$ given by 
Eq.(\ref{e11}), such that Eq.(\ref{e10}) 
is satisfied. On the boundary, Eq.(\ref{e10}) yields 
\begin{eqnarray} \label{e12} 
\int G(x(j),x(s)) \Phi(s) ds  \ \ = \ \ 0 
\ \ \ \mbox{[BIM equation]} 
\end{eqnarray} 
Thus, having an eigenstate $\Psi(x)$ implies 
that the kernel $G(x(j),x(s))$ has an eigenvector 
$\Phi(s)$ that corresponds to a zero eigenvalue.
Fig.2 shows an example 
of a boundary charge density $\Phi(s)$ that was found 
via the BIM equation (for more details, see next section). 
The converse 
is also true: Once a non-trivial charge density is 
found that satisfies Eq.(\ref{e12}), the associated
eigenstate can be constructed using Eq.(\ref{e10}).
We discuss this construction issue in more details 
in the next section.  

For numerical purposes, it is convenient to use the 
discretized version Eq.(\ref{e7}) of the above formula. 
The BIM equation can then be 
written as the matrix equation $\mbf{A}\Phi=0$, 
where $\mbf{A}_{js} = G(x(j),x(s))$. The gauge 
term ${\cal C}(x,x')$ allows some freedom in the
determination of the Green function.
Using the Neumann Bessel function $Y_0(k|x-x'|)$ 
for the Green function, one obtains the 
matrix $\mbf{A}_{js}$ as defined 
by Eq.(\ref{e5}) with~(\ref{e3}). 
Another possibility is to use the Hankel Bessel 
function $H_0(k|x-x'|)$. Accordingly, we 
will distinguish between the Y0-BIM version 
and the H0-BIM version. We shall later discuss 
the numerical implication of using the 
complex $H(k|x-x'|)$ rather than the real $Y(k|x-x'|)$. 
 
The primitive BIM uses Eq.(\ref{e12}) literally. 
However, this version of the BIM is not the one that 
is generally favored because  $G(x(j),x(s))$ 
is singular for $x(j)\rightarrow x(s)$, 
leading to some difficulty in determining the diagonal 
matrix elements of $\mbf{A}_{js}$. 
Therefore, other versions of the BIM 
have become popular (see Appendices B,C).

\subsection{The GFM} 
 
The GFM is a different strategy to obtain the 
charge density $\Phi(s)$. Rather than 
using the BIM equation Eq.~(\ref{e12}) 
or one of its variants, a gauge freedom argument 
is invoked in order to introduce 
a new type of equation (Eq.(\ref{e13}) below). 
It is clear that Eq.(\ref{e12}) should be 
valid for {\em any} choice of gauge. 
In other words, Eq.(\ref{e12}) should be 
satisfied for any Green function (Eq.(\ref{e8})), 
whatever the choice of ${\cal C}(x,x')$. 
Thus, for a given ${\cal C}(x,x')$, 
the charge density $\Phi(s)$ must satisfy 
the equation 
\begin{eqnarray} \label{e13} 
\int {\cal C}(x(j),x(s)) \Phi(s) ds  \ \ = \ \ 0 
\ \ \ \mbox{[GFM equation]} 
\end{eqnarray} 
For example, we may take ${\cal C}(x,x') = J_0(k|x-x'|)$, 
and we shall refer to this version of GFM as J0-GFM. 
For numerical purposes, it is once again convenient to 
discretize the integral expression, which can then be
written as the matrix equation $\mbf{A}\Phi=0$.

The kernel $\mbf{A}_{js} = J_0(k|x-x'|)$ 
of the J0-GFM is non-singular, and very well-behaved. 
Thus, the J0-GFM method, unlike the Y0-BIM, provides 
an extremely convenient way of obtaining the eigenvalues 
of the Helmholtz equation. Fig.2 shows an example 
of a charge density that was found via the J0-GFM equation 
(for more details see next section). 
The result is indistinguishable from the charge density
generated by the traditional H1-BIM. 
[We note however that the J0-GFM method has certain 
numerical limitations that we are going to discuss later]. 
Once the eigenvector $\Phi(s)$ is found via the GFM equation, 
we can proceed as with the traditional BIM, and construct 
the wavefunction $\Psi(x)$ using Eq.(\ref{e7}).

\section{Constructing the wavefunction} 
\label{III} 

In this section we explain how a wavefunction $\Psi(x)$ 
is constructed for a given $k$. It is assumed that $k$ is an 
eigenvalue. The (numerical) question how to determine the 
eigenvalues $k=k_n$ is differed to Section~4.

\subsection{Green function method (Eq.~(\ref{e7}))} 
\label{III_A} 
 
Both the BIM and GFM make use of Eq.(\ref{e7}) in order
to construct the wavefunction. In order to find 
the charge vector $\Phi_s$ the BIM equation (Eq.(\ref{e12})) 
and the GFM equation (Eq.(\ref{e13})) 
are written as the matrix equation \mbox{$\mbf{A}\Phi=0$}. 
The only difference between the two 
is in the expression for $\mbf{A}$. 
Given $k$, one performs the SVD 
of the matrix~$\mbf{A}$. Fig.4 displays an example 
of the output of such a SVD procedure. 
One then finds the right eigenvector $\Phi$ that 
corresponds to the {\em smallest} singular value.

Once the charge vector $\Phi_s$ has been determined, 
as in the example of Fig.2, one can construct 
the wavefunction using Eq.(\ref{e7}). 
For the Green function Eq.(\ref{e8}), it is 
most natural to use the simplest gauge (${\cal C}=0$). 
If $k$ is known to be an eigenvalue, 
then any gauge should give the same result, 
and in particular, the wavefunction
associated with any complex part of the Green 
function (such as that of the Hankel function) 
should vanish. The outcome of the Green function method 
is illustrated in Fig.3.

It is natural to ask how the constructed wavefunction $\Psi(x)$ 
look like outside of the boundary. The answer turns out to be 
$\Psi(x)=0$. For completeness, we give a proof 
of this statement. Let us define an extended function 
$\Psi_{\tbox{ex}}(x)$ such that 
$\Psi_{\tbox{ex}}(x)=\Psi(x)$ inside 
and $\Psi_{\tbox{ex}}(x)=0$ outside of the boundary. 
We would like to show that $\Psi(x)$ as defined 
by Eq.(\ref{e10}) is also equal to $\Psi_{\tbox{ex}}(x)$ 
outside of the boundary. 
It is clear that $\Psi(x)$ is a solution of the GPL equation 
by construction [see discussion following Eq.(\ref{e9})]. 
In the next paragraph, we are going to argue 
that $\Psi_{\tbox{ex}}(x)$ is a solution of 
the {\em same} GPL equation. It follows that 
the difference $R(x)=\Psi(x)-\Psi_{\tbox{ex}}(x)$ 
is a solution of Helmholtz equation in free space. 
From the definition of $\Psi_{\tbox{ex}}(x)$, 
we have $R(x)=0$ in the interior region, 
which implies by the unique continuation property 
that $R(x)=0$ over all space. 
 
The proof that $\Psi_{\tbox{ex}}(x)$ is a solution of 
the GPL equation with a charge density given by 
Eq.(\ref{e11}) goes as follows: By construction, 
$\Psi_{\tbox{ex}}(x)$ satisfies the GPL equation inside 
as well as outside of the boundary. All we have to show 
is that it also satisfies the GPL equation across 
the boundary. The latter statement is most easily 
established by invoking Gauss' law. This approach 
is valid because at short distances, $G(x,x')$ coincides 
with the electrostatic Green function. 
Thus, the gradient of $\Psi_{\tbox{ex}}(x)$ corresponds, 
up to a sign, to the electric field. Gauss' law implies 
that the electric field should have a discontinuity 
equal to the charge density $\Phi(s)$. 
Indeed, $\Psi_{\tbox{ex}}(x)$ is consistent with 
this requirement.

\subsection{Decomposition method (Eq.~(\ref{e6}))}
\label{IIIB}

The other procedure to construct the wavefunction
is to use the DEM Eq.(\ref{e6}). The idea is to regard $F_j(x;k)$
as a basis for the expansion:
\begin{eqnarray} \label{e16_0}
\Psi(x) \approx \sum_j  {\mathsf C}_j F_j(x;k)
\end{eqnarray}
Any such superposition at fixed $k$ is a solution of
the Helmholtz equation within the interior region. Thus, in order
to satisfy the Dirichlet boundary conditions, one looks for a vector
${\mathsf C}$ of expansion coefficients that 
satisfy ${\mathsf C} \mbf{A} = 0$.
It turns out that the direct numerical implementation of 
this simple idea is a complicated issue (see discussion of 
the null space problem later in this section).

Any set of $F_j(x;k)$ which are solutions of the Helmholtz equation 
may be used for the DEM. However, it should be remembered
that computationally not all bases are equivalent.
For instance, the $Y_0$ basis defined by Eq.(\ref{e3})), which might
appear to be the best choice as a DEM basis due to its association
with the BIM, does not give the best numerical results
when compared against other options.
In particular, it turns out that the PWDM is typically much more
effective (recall that the PWDM is a special case of
the DEM, corresponding to the choice Eq.(\ref{e2}) of basis functions.)
Finally we note that the set of $J_0$s of Eq.(\ref{e4})
can not be regarded  as a mathematically legitimate basis
for a DEM. This latter point will be explained in Section~5.

For the $Y_0$ basis the BIM and the DEM lead to the same equation.
It is only the mathematical interpretations that is different. 
Within the DEM, one regards the $Y_0$ as basis functions to be used 
in an expansion, while the same $Y_0$ in the BIM context serves  
as the Green function. In the context of DEM, one may be bothered 
by the singular nature of the $Y_0$ functions: The constructed wavefunction 
should be zero on the boundary.  Mathematically this is achieved 
in the $N\rightarrow\infty$ limit, so the $Y_0$ basis is a valid choice.
But in an actual numerical implementation, the wavefunction so constructed 
will always have singularities on the boundary. 
One possible remedy consists of enforcing the boundary conditions 
on intermediate boundary points, or on points that lie 
on an outer boundary. Alternatively, one may replace the 
bare $Y_0$ basis by smooth  superpositions of $Y_0$ functions (see Appendix C).

In Fig.2, we display an example of a numerically determined
${\mathsf C}$ (using the PWDM basis) for one of the Pond eigenstates,
while in Fig.~3 we illustrate the constructed wavefunction.
Unlike the Green function construction,
the DEM/PWDM constructed wavefunction does not
vanish outside of the boundary. Actually, it is
quite the opposite: Typically the DEM/PWDM wavefunction
becomes exponentially large as we go further away from 
the boundary. Whenever this behaviors occurs, 
it constitutes an indication
of the {\em evanescent} nature of the wavefunction.
Namely, in such cases, the wavefunction acquires
sub-wavelength features in order to
accommodate the boundary. This requires
exponential behavior (negative kinetic energy)
in the transverse space direction, in order to keep 
the total energy fixed.

\subsection{Normalization of the wavefunction}

A standard SVD procedure generates vectors $\Phi_s$ and ${\mathsf C}_j$ 
that are normalized in the sense $\sum_s |\Phi_s|^2 = 1$ and 
$\sum_j |{\mathsf C}_j|^2 = 1$.  Therefore, the constructed $\Psi_r$  
is not properly normalized within the interior region.
Adopting the usual philosophy of boundary methods, 
the problem of calculating the $\Psi_r$ normalization 
is reduced to that of evaluating a boundary integral, 
namely~\cite{berry1,boasman} 
\begin{eqnarray} \label{e15} 
\int\int |\Psi(x)|^2 dx = 
\frac{1}{2k^2} \oint |\Phi(s)|^2 (n(s){\cdot}x(s))ds 
\end{eqnarray} 
For the BIM, by discretizing of Eq.~(\ref{e15}) 
we obtain the following numerical 
expression for the normalization factor: 
\begin{eqnarray} \label{e15a} 
\| \Psi_r \| \ \ = \ \ \frac{1}{\Delta s} 
\sum_s \mbf{w}_s (\Phi_s)^2 
\ \ = \ \ \Phi^{\dag} \mbf{W} \Phi. 
\end{eqnarray} 
Here $\mbf{W}=(1/\Delta s)\mbox{diag}(\mbf{w}_s)$ 
is a diagonal matrix, and the weight factor 
$\mbf{w}_s$ is defined in Table~1. 
As for the DEM, by using the derivative of Eq.(\ref{e16_0}) 
in  Eq.(\ref{e11}) and substituting into Eq.(\ref{e15}), we get: 
\begin{eqnarray} \label{e16a} 
\| \Psi_r \| \ &=& \ \Delta s \sum_s \mbf{w}_s 
\left(\sum_j {\mathsf C}_j \mbf{D}_{js}\right)^2 
\\ \label{e16aa} 
\ &=&  \ \sum_{ij} {\mathsf C}_i \mbf{B}_{ij} {\mathsf C}_j 
\ = \ {\mathsf C}\mbf{B}{\mathsf C}^{\dag} 
\end{eqnarray} 
The definitions of $D_{js}$ and of the metric $B_{ij}$ 
can be found in Table 1. 

The normalization $\| \Psi_r \|$ can be calculated 
using the metric $\mbf{B}_{ij}$. This method looks quite 
elegant, but it turns out not to be very effective 
numerically. Consider Eq.(\ref{e16a}). This 
equation is quite safe computationally for two reasons: 
(i) all its terms are non-negative; 
(ii) standard summation routines order these terms 
in descending order. Now, let us look instead at 
Eq.(\ref{e16aa}). In this case the numerical calculation 
can give {\em any} result (if we go 
to large $k$). Sometimes, the answer even 
comes out to be negative! This occurs because the calculation 
involves many arbitrarily ordered terms that 
each have a different algebraic sign.

\subsection{The tension along the boundary}

The numerical wavefunction $\Psi_r$ satisfies Helmholtz 
equation in the interior region {\em by construction}. 
Thus, whether $\Psi_r$ is an actual eigenstate depends 
on its behavior along the boundary. 
In this subsection we would like to discuss 
the definition of a `tension' measure 
that estimates whether, and to what accuracy, 
the numerical $\Psi_r$ satisfies the boundary conditions.

For the case of the DEM, following \cite{heller2}, 
the tension is defined as the boundary integral 
\begin{eqnarray} \label{e16b} 
\| \Psi_s \|  \ &=& \ \Delta s \sum_s 
\left(\sum_j {\mathsf C}_j \mbf{A}_{js}\right)^2  
\ = \ \sum_{ij} {\mathsf C}_i \mbf{T}_{ij} {\mathsf C}_j 
\ = \ {\mathsf C}\mbf{T}{\mathsf C}^{\dag} 
\end{eqnarray} 
The standard practice to date for the 
tension calculation has been to 
use a denser set of boundary points 
located between the $x_s$ points. 
However, our experience (see also \cite{barnett}) 
is that the tension estimate obtained from 
the initial set of points is just as effective. 
This is demonstrated in Fig.3d. 
Therefore, we routinely rely on the same set 
of boundary points to determine the tension.

For the BIM on the other hand, the above definition 
is not practical due to the singular nature of the basis functions. 
For any finite $M$, the numerical wavefunction 
blows up at each boundary point. 
However, since the BIM wavefunction vanishes everywhere 
outside of the billiard, a numerically unambiguous definition 
of tension arises as an integral of $|\Psi(x)|^2$ along 
an outer boundary: 
\begin{eqnarray} \label{e15b} 
\| \Psi_ {{\mathsf s}}\| \ = \ 
\Delta s \ \sum_{{\mathsf s}} (\Psi_{{\mathsf s}})^2 
\ = \ \Delta s \ \sum_{{\mathsf s}} 
\left(\sum_s\mbf{G}_{{\mathsf s}s}\Phi_s \right)^2 
\end{eqnarray} 
By outer boundary (see Fig.1), we mean the set of 
external points (${\mathsf s}$~points, as opposed to $s$~points 
for the true boundary) that have a fixed 
transverse distance $\Delta L$ from the the true boundary. 
The distance $\Delta L$ between the boundary and the outer boundary 
should be small on any classical scale but large compared with $ds$, 
in order for the tension to be independent of the 
choice of $\Delta L$. See Fig.5c for a numerical demonstration.

\subsection{The PWDM and the null space problem}

One may think that ${\mathsf C}$ could be found
simply by computing the (left) eigenvector
that corresponds to the smallest singular
value of $\mbf{A}_{js}$. Numerically this definition 
is hard to implement. This difficulty can be explained 
by looking at the behavior of the singular values 
of $\mbf{A}_{js}$ for the PWDM basis.
Fig.4 gives some examples of singular values
resulting from the SVD of the $\mbf{A}_{js}$ matrix.
In the case of the PWDM, as $k$ become large, one observes 
that the the singular values separate into two groups: 
rather than having one distinctly smaller singular 
value, we obtain a whole set of them. 
Accordingly, we can define a numerical `null space' 
of the $\mbf{A}_{js}$ matrix.

The interpretation of this null space is quite clear. 
It is well known \cite{dietz,uzy} that it is not efficient 
to include much more than $N_{sc}$ 
plane waves in the basis set $F_j(x;k)$, where 
\begin{eqnarray} \label{e16c} 
N_{sc} \ = \ \frac{1}{\pi} kL 
\end{eqnarray} 
and $L$ is the perimeter of the billiard. 
The reason for this ineffectiveness is 
that $k_i$ and $k_j$ cannot 
be distinguished numerically within the interior 
region unless \mbox{$|k_i-k_j| L > 1$}. 
In order to obtain the semiclassical result (\ref{e16c}), 
the total phase space area ($L\times(2mv)$) 
of the boundary Poincar\'e section is divided by the size
of Planck cell ($2\pi\hbar$). If we use
$N>N_{sc}$ plane waves, then we can create
{\em wavefunctions that are nearly zero in the interior},
and become large only as we go far away from the center\cite{berry2}.
It is clear that the SVD can be used to determine
the $N-N_{sc}$ null space of these evanescent states.
Whenever $k$ is an eigenvalue, this null space includes,
in addition to the evanescent waves, the single eigenvector
that constitutes an eigenstate of the Helmholtz equation.
The problem is to distinguish this eigenvector from the other
vectors in the null-space.

The basic difference between the eigenvector that leads 
to an eigenstate (which exists if $k=k_n$), and the 
other vectors in the null space is related to the normalization. 
As discussed before, a standard SVD procedure generates 
vectors  ${\mathsf C}_j$ that are normalized 
in the sense $\sum_j |{\mathsf C}_j|^2 = 1$. 
Therefore, the  $\Psi_r$ of Eq.(\ref{e6}) is not properly normalized 
within the interior region. Normalizing the wavefunction 
has the effect of magnifying the evanescent solutions 
in the interior as well as on the boundary, 
while the eigenfunction (if it exists) remains 
small on the boundary. In appendix D, we give a detailed 
explanation of the numerical procedure for 
finding ${\mathsf C}_j$ that can be derived from the above observation.

\section{The quantization measure} 
\label{IV}

Once we have constructed the wavefunction at a given $k$, the next step 
is to determine whether $\Psi$ is an eigenstate. As we will explain below, 
our choice of measures reduces to finding the minima of one of: 
\begin{eqnarray} \label{e28a} 
S(k) \ &=& \ \mbox{tension} 
\\ \label{e28b} 
S(k) \ &=& \ \mbox{smallest singular value} 
\\ \label{e28c} 
S(k) \ &=& \ \mbox{determinant} 
\end{eqnarray} 
We call $S(k)$ the quantization measure.
Below we give further explanation of the above definitions.

It is clear that the most natural quantization measure is the tension. 
If a properly normalized wavefunction has ``zero tension" on the boundary, 
it means that the corresponding $k$ is an eigenvalue. 
The normalization issue is further discussed in Appendix~D.  
The question that arises is whether we can use a numerically simpler  
measure, and what price we pay for doing so.

The BIM Eq.~(\ref{e12}) and the GFM Eq.~(\ref{e13}) can both be written as 
$\mbf{A}\Phi = 0$, with the appropriate choice of  $\mbf{A}$. 
Thus, if $k$ is an eigenvalue, $\mbf{A}$ should have a singular value 
that tends to $0$ as $N$ increases. The determinant of $\mbf{A}$ 
is defined as the product of all the singular values, and therefore 
it should vanish whenever one of the singular values does. 
Using the {\em GFM-DEM duality} which is discussed in Section~5, 
it is clear that for the DEM (and for the PWDM in particular) 
the determinant of $\mbf{A}$ vanishes whenever $k$ is an eigenvalue. 
It is important to realize that in the latter argumentation, 
{\em we do not rely on inside-outside duality} \cite{uzy} considerations, 
but rather on the much simpler {\em GFM-DEM duality}.

Thus, a low tension must be correlated with
having a vanishingly small singular value or determinant. 
The converse is not true: It is well known that 
SVD based quantization measures may lead to spurious 
minima (see \cite{backer} and references therein). 
Therefore SVD based procedure for finding eigenvalues 
requires a post-selection procedure whose aim  
is to distinguish true zeros from fake ones.

It is important to realize that neither the traditional implementation 
of PWDM, nor that of the BIM should be considered to be 'package deals'. 
For example, the BIM could be used with the 
tension as a measure (defined in the next section), rather than looking 
for minima of the the singular values. 
Similarly, the usual Heller method of PWDM implementation (see Appendix~D) 
could be replaced by a search over determinant values.

\subsection{The tension as a quantization measure}

The tension is a robust measure of quantization.
Fig.5 displays some examples of the corresponding 
$S(k)$ plots. The PWDM minima 
are typically much sharper than their BIM equivalents. 
Zooming over a PWDM minimum (Fig.5d) reveals 
some amount of roughness. 
This feature is actually helpful, because it 
gives an indication of and control over the accuracy 
of the numerics. We interpret the roughness of the 
PWDM minimum as a reflection for the 
existence of a null-space. In the same spirit, 
the smoothness of the BIM minima can be regarded 
as an indication that better accuracy can be 
obtained by making $N$ larger. We discuss these issues 
below.
 
The tension provides a common measure that 
may be used to monitor improvements, 
as well as to compare the success of the different methods. 
Naturally, the first issue to discuss is 
the dependence of the tension on the 
size $N$ of the basis set (see Fig.6). 
For the BIM, the tension becomes better as $N$ grows, 
and disregarding the computer hardware, 
there is no reason to suspect that there is an inherent 
limitation on the accuracy. The situation is different for the PWDM. 
Here, taking $N$ much larger than $N_{sc}$ is not effective. 
In practice, the method reaches a limiting accuracy, 
which, taking into account present hardware limitations, 
is still very good compared with that of the BIM. 
 
From Fig.6, it is also clear that the tension of the PWDM becomes 
much better as $k$ becomes larger. This is expected on the 
basis of the following semiclassical reasoning: larger uncertainties 
in $k$ result from confining a particle 
to a smaller box (taking a smaller box for a given $k$ 
is equivalent to making $k$ smaller for a given box size). 
Thus, it is more difficult to build a wavefunction with a precise 
value of $|k_j|=k$ for low lying eigenstates. 
On the other hand, the BIM does not seem to be sensitive 
to the value of $k$.

\subsection{The tension as an indication for the global error}

The tension can be regarded as a measure 
of the {\em local error} in the 
determination of the eigenfunction. 
The tension is local in the sense that it pertains only 
to points along boundary. We can also define 
a measure for the {\em global error}, 
that is the error which is associated with all 
the interior points:
\begin{eqnarray} \label{e_error} 
(\Delta \Psi)^2 \ \ = \ \ 
\langle |\Psi_r - \Psi_{\tbox{exct}}(X_r)|^2 \rangle 
\end{eqnarray} 
Here $\Psi_{\tbox{exct}}(x)$ is the numerically exact 
wavefunction. The average is taken over the 
set $X_r$ of selected points inside of the boundary. 
Fig.7 gives an example for the variation of the 
error along the cross section line of Fig.1. 
In order to eliminate a possible bias due to a 
global normalization error, we renormalized 
the inexact wavefunction so that 
$\Psi_r = \Psi_{\tbox{exct}}(X_r)$ at a randomly selected 
point $X=X_0$. In retrospect, we realized that 
such an error did not significantly affect the result. 
However, we still chose to be on the 
safe side, and we adopted this procedure routinely. 
  
It is natural to expect the average error 
$(\Delta \Psi)^2$ to be correlated with 
the tension. In other words, if 
$|\Psi_r - \Psi_{\tbox{exct}}(X_r)|$ is small on the boundary, 
then one may expect it to be small in the interior. 
The degree of such correlation is important 
for practical reasons. Moreover, we have introduced two different versions 
of tension definitions, one for each of the PWDM and the BIM. 
It is not a-priori clear that the above correlation is independent 
of the choice of numerical method. 
In Fig.8, we study this issue by plotting $(\Delta \Psi)^2$ 
against the tension for the BIM and the PWDM. 
In the case of the PWDM, the error saturates below a critical 
tension. After this point, further improvements on the boundary 
do not seem to affect the bulk of the eigenstate. 
It is not clear from the numerics whether or not the BIM 
saturates. What is clear however is that 
the BIM does a poorer job at reproducing 
the wavefunction inside of the boundary than the PWDM with 
the same tension. 
 
The saturation of the error well inside the billiard 
can be explained as a manifestation of the fact that 
the wavefunction there is not very sensitive to 
sub-wavelength roughness of the boundary: 
If $N$ is reasonably large, the numerical wavefunction 
vanishes on a nodal line that almost coincides with the 
true (pre-defined) boundary. Increasing $N$ further 
affects the sub-wavelength features of the 
(distance) difference between that nodal-line and 
the true boundary. This distance difference is 
important for the tension, but barely
affects the wavefunction well inside the billiard.


\subsection{The determinant as a quantization measure}

The tension is the natural choice for a quantization measure. 
However, from a numerical point of view, it is much more convenient 
and time effective to compute the singular values of $\mbf{A}$, 
without having to find the eigenvectors for each $k$ value, 
and without having to compute the wavefunction 
along the boundary (for tension calculation).

The smallest singular value is traditionally used 
as a quantization measure for the BIM. 
From Fig.7, it is quite clear that for the BIM 
one of the singular values is significantly smaller 
than the others, so that the eigenstate 
is unambiguously determined by this method. 
Is it possible to use the same approach, with comparable success, 
for the PWDM?  We have already determined that looking at the 
smallest singular values is not very meaningful numerically. 
For $N > N_{sc}$, there exists a large null-space of evanescent 
states for any $k$.  The metric method (Appendix D) 
is not a practical solution to this problem 
since we want to gain numerical efficiency (if efficiency is not 
the issue then it is better to use the tension as a quantization measure).

One simple way to improve the numerical stability 
is to use the determinant rather than the smallest 
singular value as a quantization measure:  
Each time that $k=k_n$, the null space should include one more 
`dimension'. Therefore, the determinant, rather than the 
smallest singular value, becomes the reasonable quantity 
to look at. Thus, from numerical point of view (\ref{e28b}) 
should be superior compared with (\ref{e28a}).

Fig.9 illustrates how the determinant can be used in practice 
as a quantization measure. As a general rule, as is the case 
for the tension, the PWDM/GFM minima are sharper than the BIM 
ones. On the one hand, this extra sharpness can be regarded as 
an advantage, because it leads to a better resolution of the 
eigenvalue spectrum. However, more computer time is needed in order 
to find these minima.  The BIM minima are broader, 
and therefore digging algorithms that search for local minima 
become extremely effective.

In the case of the traditional BIM, using a larger $N$ leads to 
a better resolution of the local minima, as expected. 
The traditional H-BIM uses the complex Hankel Bessel 
function as its Green function. One may wonder why 
the real Neumann function could not be 
used instead. A-priori, there is no reason to 
insist on Hankel choice. However, it seems that 
with Neumann choice the numerics are not very stable:  
The locations of the local minima vary on a $k$ range which is 
large compared to their $k$ width. 
Because of this problem, search routines 
relying on the Y-BIM may yield misleading values for the 
error in the $k_n$ determination.
Thus, the numerical stability of the H-BIM can be attributed 
to the fact that the BIM equation $\mbf{A}\Phi=0$ becomes 
complex. Its real part is just the Y-BIM equation, 
while its imaginary part is the J-GFM equation. 
Thus one may say that the H-BIM benefits from combining 
the Y-BIM with the J-GFM.

Is it practical to use the Fredholm determinant 
as a quantization measure also in the GFM/PWDM case? 
Here we observe that the null-space problem is reflected 
in the stability of the determinant calculation.
It is useful to characterize the numerics 
using the discretization parameter~$b$:
\begin{eqnarray} \label{e_bdef}
b \ \ = \ \ \frac{N}{N_{sc}} \ \
=\Big|_{\tbox{$M$=$N$}} \ \ \frac{\lambda/2}{\Delta s}
\end{eqnarray}
The last equality holds if we take $M=N$,
leading to the interpretation of $b$ as the
number of boundary points per half De-Broglie wavelength.
If $b<1$ the null space problem does not exist,
and we can define ${\mathsf C}$ as the
(left) eigenvector that corresponds to the smallest
singular value of $\mbf{A}_{js}$.
Of course, we want to push PWDM to the limit,
and therefore in practice we always take $b>1$.

The natural question is whether choosing a
very large $b$ is numerically useful.
Our numerical experience is that for $1<b<1.8$ 
we get nice minima, which actually look much 
sharper than the BIM ones (see Fig.~9). 
As we try to increase $b$ in order to improve accuracy, 
the numerics loose stability (what we mean by 
instability is demonstrated in Fig.~9e). 
The same phenomenon occurs with J0-GFM, 
which has somewhat larger tendency for instability. 
This is apparently because the J0-GFM 
is involved with a larger null-space (see Fig.~4).

Thus we conclude that taking $b>1$ does improve the
numerics, while $b \gg 1$ generally leads to
instabilities that should be avoided.
The optimal choice of $b$ depends
on the details of the implementation
and on the computer hardware. It should be clear that if the
numerical accuracy were unlimited, 
then the $b\rightarrow\infty$ limit would lead 
to a numerically exact solution in cases
where the wavefunctions may be written 
as superpositions of plane waves.  
This is not always possible \cite{boris}.
Note however that evanescent features  
of the wavefunction can be reconstructed 
by a suitable superposition of plane waves \cite{berry2}.

\section{The duality of the GFM and the DEM} 
\label{V} 

In addition to providing a boundary method of its own, 
the GFM also serves to bridge the gap between 
the BIM and the DEM. Consider the version of the GFM 
that is based on the choice ${\cal C}(x,x') = F_j(x;k)$, 
where the $F_j$ are solutions of the Helmholtz equation 
in free space (with neither singularities nor cuts). 
With this choice, we immediately realize that 
the DEM and the GFM are dual methods: 
\begin{eqnarray} 
\label{e14a} 
\mbf{A} \Phi=0  \hspace*{1.5cm} & \mbox{[GFM equation]} \\ 
\label{e14b} 
{\mathsf C} \mbf{A}=0  \hspace*{1.5cm}  & \mbox{[DEM equation]} 
\end{eqnarray} 
The only difference lies in whether one looks 
for the {\em left} or the {\em right} eigenvector. 
This point is numerically demonstrated in Fig.2. 
  
The PWDM version of the DEM also satisfies this duality. 
In this special case, a somewhat more elegant version 
of the above argument is as follows: 
Consider the version of the GFM that is based on the choice 
\mbox{${\cal C}(x,x') = \exp(ik n_j\cdot(x-x'))$}, 
where $n_j$ is a unit vector in a given direction. 
We can take $N$ different choices of $n_j$, 
thus obtaining the matrix equation $\mbf{A} \Phi = 0$ with 
\begin{eqnarray} \label{e14} 
\mbf{A}_{js} \ \ = \ \ 
\exp(ik n_j \cdot x_s) 
\end{eqnarray} 
An equivalent matrix equation is found by 
multiplying each equation by $\exp(ik\phi_j)$, 
where  $\phi_j$ are random phases. We can then 
take the real part of these equations, thus 
obtaining a set of equations that involves the same 
matrix $\mbf{A}$ as that of PWDM, 
namely (\ref{e5}) with the basis defined by (\ref{e2}).

The duality of the PWDM and the GFM is very important from 
a mathematical point of view. The mathematical 
foundations of the PWDM are quite shaky. 
It is clear that PWDM is well-established mathematically 
if a strict `inside-outside duality' (IOD) \cite{uzy} is satisfied.
%
%
In this case, the wavefunction can be extended to the
whole plane so that the boundary of the billiard can be 
regarded as a {\em nodal line} of some plane-wave superposition. 
Obviously, this is rarely possible \cite{berry2}. 
Therefore, one may wonder whether 
we indeed have $\det(\mbf{A})=0$ whenever $k=k_n$. 
Using the duality Eq.(\ref{e14a}), it becomes obvious 
that the Fredholm determinant indeed vanishes at the eigenstates, 
even in the absence of an exact IOD. 
 
It is quite clear from the first paragraph 
of this section that any expansion method can 
be associated with a corresponding GFM. 
Whenever the left eigenvector 
is used with the expansion method, 
the right eigenvector can be used with the GFM. 
We have already demonstrated this point in Fig.2. 
Is it possible to make the inverse statement? 
Do we have a well defined expansion method 
associated with any GFM?  The answer is negative. 
We discuss this issue in the rest of 
this section, and it can be skipped at first reading.  

For the following, it is convenient to 
consider Eq.(\ref{e6}) as $N\rightarrow\infty$. 
Subsequently, we are going to talk about whether this 
limit is meaningful. 
In the case of the usual PWDM, 
the $N\rightarrow\infty$ limit of Eq.(\ref{e6}) 
can be written as 
\begin{eqnarray} \label{e16_1} 
\Psi(x) = \int_0^{2\pi} 
C(\theta)d\theta 
\ \exp(ik n(\theta)\cdot x) 
\end{eqnarray} 
Similarly, in the case of the $J_0$ decomposition, 
using the basis functions of Eq.(\ref{e4}), 
we can write in complete analogy: 
\begin{eqnarray} \label{e16_2} 
\Psi(x) = \oint 
\Phi(s)ds \ J_0(k|x-x(s)|) 
\end{eqnarray} 
In writing Eq.(\ref{e16_2}) we have used the fact 
that $J_0(x(j)-x(s))$ is a symmetric kernel, 
and therefore we could make the substitution ${\mathsf C}=\Phi$.

Eq.(\ref{e16_2}) looks at first sight 
like an innocent variation of Eq.(\ref{e16_1}). 
The Bessel function $J_0$ is just a superposition 
of plane waves, and therefore one may 
possess the (incorrect) idea that there is a simple 
way to go from Eq.(\ref{e16_2}) to Eq.(\ref{e16_1}). 
If we expand each $J_0$ in Eq.(\ref{e16_2}) 
in plane waves, and re-arrange the expression 
in order to identify the PWDM coefficients, 
we end up with the relation 
\begin{eqnarray} \label{e18} 
{\mathsf C}(\theta)  = 
\int \mbox{e}^{- ik n(\theta){\cdot}x(s)} 
\ \Phi(s) ds 
\end{eqnarray} 
This relation implies the trivial result 
${\mathsf C}(\theta)=0$ due to gauge freedom 
[see discussion of Eq.(\ref{e14})]. 
Hence we conclude that the constructed 
wavefunction is $\Psi(x) \equiv 0$ in the 
$N\rightarrow\infty$ limit! 
  
Having $\Psi(x) \equiv 0$ from Eq.(\ref{e16_2}) 
could have been anticipated using a simpler 
argument: We know that Eq.(\ref{e10}) should hold 
for {\em any} gauge choice. 
This gauge freedom implies that we have 
\begin{eqnarray} \label{e17} 
\oint {\cal C}(x,x(s)) \Phi(s) ds  \ \ = \ \ 0 
\end{eqnarray} 
The above equation should hold for 
any $x$ inside 
as well as on the boundary. 
Furthermore, the left hand side 
of (\ref{e17}) is manifestly a solution 
of Helmholtz equation in free space, 
and it follows from the unique continuation 
property that Eq.(\ref{e17}) holds also 
for points outside of the boundary. 
Having $\Psi(x)=0$ as a result of the 
integration in Eq.(\ref{e16_2}) 
is just a particular case of Eq.(\ref{e17}). 
 
In spite of the observation that Eq.(\ref{e16_2}) 
yields $\Psi(x) \equiv 0$ in the $N\rightarrow\infty$ 
limit, the vector $\Psi_r$ is non-zero numerically 
for any finite $N$.  In fact, 
after proper re-normalization, $\Psi_r$ becomes a quite good 
approximation to the wavefunction (see Fig.3c and Fig.3e). 
Sometimes the result so obtained is 
even better than the one which is found 
via the traditional BIM Eq.~(\ref{e7}). 
As strange as it sounds, this success is entirely 
due to the fact that we are using finite $N$. 
  

\section{Conclusion}

The BIM and the DEM were written as different faces of a unified boundary
procedure comprising four steps. In the process of doing so, yet a third 
boundary method was derived as part of the same framework, the GFM. 
The DEM and the GFM are strongly related, as they are respectively the 
left and right eigenvectors of the same Fredholm matrix. We think that 
the presented approach opens the way towards a controlled fusion 
of the BIM and the DEM into a more powerful numerical procedure. 

The unified treatment of quantization measures allowed us to compare 
the efficiency of the various methods, and to analyze both the local 
and the global errors in the numerically determined wavefunction. 
In particular, a numerically valid definition of the BIM tension was given, 
and was found to possess smooth minima at the eigenstates. 
Using the tension as a quantization measure is one possible way 
to avoid some problems \cite{backer} that are encountered in the traditional 
implementation of the BIM. 

\appendix

\section{The BIM for the scattering problem} 
 
The solution of the Helmholtz equation for the scattering problem 
is just another variation of the BIM. Consider a boundary, one that in 
general may be composed of several disconnected pieces. The incident 
wave $\Psi_{\tbox{incident}}(x)$ is a solution of Helmholtz 
equation in free space. Formally, $\Psi_{\tbox{incident}}(x)$ includes both 
the ingoing and the outgoing wave components. We look for a solution 
$\Psi(x)$ that has the same ingoing 
component as $\Psi_{\tbox{incident}}(x)$, 
and that satisfies $\Psi(x)=0$ on the boundary. 
Such solution can be written as a sum 
of the incident wave and a scattered wave, 
and hence must be of the form 
\begin{eqnarray} \label{eb1} 
\Psi(x) = \Psi_{\tbox{incident}}(x) 
+ \oint G(x,x(s')) \Phi(s') ds 
\end{eqnarray} 
Equation (\ref{eb1}) is a variation of Eq.(\ref{e10}). 
Note that the Green function should 
satisfy outgoing boundary conditions in order to 
yield the desired solution. 
The charge density $\Phi(s)$ is fixed by 
the requirement that $\Psi(x)=0$, which leads to the boundary equation 
\begin{eqnarray} \label{eb2} 
\oint G(x(s),x(s')) \Phi(s') ds = - \Psi_{\tbox{incident}}(s) 
\end{eqnarray} 
This inhomogeneous equation is a straightforward 
generalization of Eq.(\ref{e12}). A discretized 
version of it was used  in Ref.\cite{lupo} in order 
to obtain numerical solutions of the Helmholtz equation 
for some scattering problems.

The derivation of Eq.(\ref{eb2}) in Ref.\cite{lupo} 
is much more complicated than ours, and involves the use 
of Lippmann-Schwinger equation. 
The boundary is represented by a large delta-potential $V$, 
and the limit $V\rightarrow\infty$ is taken. 
Using this procedure, the charge density $\Phi(s)$ can 
be obtained as the $V\rightarrow\infty$ limit of $V\Psi(x(s))$. 
Note the correctness of the physical units. Namely, 
$[{\cal H}][\Psi]=[\rho]$ and therefore $[V][\Psi]=[\Phi]$. 
Note also that the wavefunction is in general non-zero 
in both sides of the boundary. Therefore the charge density 
$\Phi(s)$ is equal to the {\em difference} between the 
normal derivatives on both sides of the boundary. 
The simplest way to derive the relation between $\Phi(s)$ 
and $V\Psi(x(s))$ is to integrate the Helmholtz equation 
over an infinitesimal range across the boundary, 
as in the treatment of the 1D Shchrodinger equation 
with delta potential.

\section{Traditional BIM equation} 
 
The primitive BIM equation (Eq.(\ref{e12})) 
is based on Eq.(\ref{e10}). 
The traditional BIM is a variation of the same idea 
which avoids the boundary singularities that 
plague the more  simplistic version. 
Rather than using Eq.(\ref{e10}) directly, 
one considers its gradient, leading to 
\begin{eqnarray} \label{ea1} 
\partial_{\pm} \Psi(x(s)) \ = \ 
\oint \partial_{\pm} G(x(s),x(s')) \Phi(s') ds' 
\end{eqnarray} 
This equation is analogous to Eq.(\ref{e12}). 
We use the notation $\partial_{+}$ and $\partial_{-}$ 
in order to refer to the normal derivative 

on the interior and exterior sides of the boundary 
By definition, \mbox{$\partial_{-} \Psi(x(s)) = \Phi(s)$}, 
and from the discussion in section III 
we have \mbox{$\partial_{+} \Psi(x(s)) = 0$}. 
Adding the two equations of (\ref{ea1}), we obtain 
\begin{eqnarray} \label{ea2} 
\Phi(x(s)) \ = \ 
\oint 2\partial G(x(s),x(s')) \Phi(s') ds' 
\end{eqnarray} 
where $\partial \equiv (\partial_{+}+\partial_{-})/2$ 
is just the derivative on the boundary in the principal sense. 
Thus, in the traditional BIM, the definition 
of the Fredholm matrix is 
\begin{eqnarray} \label{ea3} 
\mbf{A}_{ss'} \ = \ \frac{1}{\Delta s}\delta_{ss'} \ 
                 - \  2\partial G(x(s),x(s')) 
\end{eqnarray} 
and the BIM equation (\ref{ea2}) is $\mbf{A}\Phi=0$. 
The kernel $\partial G(x(s),x(s'))$ is well- 
behaved, and its diagonal elements are finite 
thanks to the presence of a geometrical factor. 
Namely, if $G(x(s),x(s'))=g(k|x(s)-x(s')|)$ then 
\begin{eqnarray} 
\partial G = k\frac{n(s)\cdot(x(s)-x(s'))}{|x(s)-x(s')|} 
g'(k|x(s)-x(s')|) 
\end{eqnarray} 
If either one of the Bessel functions $H_0$ or $Y_0$ 
is used for the Green function, then the definition 
of $\mbf{A}_{ss'}$ above involves 
either $H_1$ or $Y_1$, respectively.

\section{Transformed BIM equations} 

There exists another elegant version of the BIM
\cite{klaus,klaus-long,KKR} which does not exhibit the singularities associated
with the Y0-BIM. Namely, the BIM equation is rewritten as
\begin{eqnarray}
\int \tilde{G}(x(j),\kappa)
\tilde{\Phi}(\kappa) d\kappa  \ \ = \ \ 0,
\end{eqnarray}
where the vector $\tilde{\Phi}(\kappa)$ is related by a linear transformation
to the vector $\Phi(s)$. This transformation corresponds to the selection of a 
complete basis set of boundary wavefunctions. The KKR method \cite{KKR} 
is a particular implementation which uses the spherical harmonics.
Another (more general) choice \cite{klaus,klaus-long} consists of taking
$\tilde{\Phi}(\kappa)$ as the Fourier components of $\Phi(s)$.
In the latter case $\tilde{G}$ is related to $G$ by the Fourier
transform $s\mapsto\kappa$. However, obtaining $\tilde{G}$ from $G$ is not
a simple matter for most boundary shapes.

\section{The PWDM, in practice} 
 
The mathematically clean solution for the 
null-space problem is to adopt a method based on a metric. 
As we explain in the next paragraph, 
this procedure is sensitive to cumulative numerical errors. 
A modified implementation of the metric method,  
that avoids some of the numerical problems,  
has been introduced by Barnett~\cite{barnett}. 
The other possibility is to use a very simple 
procedure which is known as Heller's method~\cite{heller2}. 
Below we discuss the latter as well.

The metric method works as follows: 
First, one finds the basis in which the 
normalization metric $\mbf{B}_{ij}$ becomes $\delta_{ij}$. 
The tension metric $\mbf{T}_{ij}$ should then be written 
in that same basis. The SVD is done on the transformed 
tension metric. In this case, the null space becomes at most 
one-dimensional (whenever $k=k_n$).  Unfortunately, this 
elegant and straightforward metric scheme does not work very well, 
due to the finite precision problems 
discussed in connection with Eq.(\ref{e16aa}). 
Furthermore, $\mbf{B}$ and $\mbf{T}$ are "squares" of $\mbf{A}$, 
which leads to a loss of numerical precision 
compared with an $\mbf{A}$-based strategy.

The most widely used $\mbf{A}$-based 
strategy is referred to as Heller's method~\cite{heller2}. 
The idea is to find ${\mathsf C}_j$ as the solution 
of the $M \geq N$ set of equations 
$\sum_j {\mathsf C}_j \mbf{A}_{js} = 0$, 
with the additional constraint 
$\sum_j {\mathsf C}_j \mbf{A}_{j0} = 1$. 
Table 1 gives the definition of $\mbf{A}_{j0}$. 
 
By constraining the wavefunction to be $\Psi(X_0)=1$ 
at a selected point $X_0$ in the interior of the billiard, 
we eliminate the problems associated with evanescent states,
as the associated (evanescent) wavefunctions no longer vanish 
on the boundary and thus, there is no longer 
a null-space problem.  As a result, quite large~$b$ 
can be used without encountering numerical instabilities. 
The only worry with this method is that $X_0$ may happen 
to be very close to a nodal line. 
In such cases, the tension will be large due to an improper 
normalization, so we will miss these eigenstates.

 
\ \\ 
 
\noindent 
{\bf Acknowledgments:} 
 
\noindent 
It is our pleasure to thank Alex Barnett, Areez Mody,
Lev Kaplan and Michael Haggerty for many useful discussions,
and Klaus Hornberger and Uzy Smilansky for their comments.
This work was funded by ITAMP and the National Science Foundation.

\ \\ 




\newpage
\centerline{\epsfig{figure=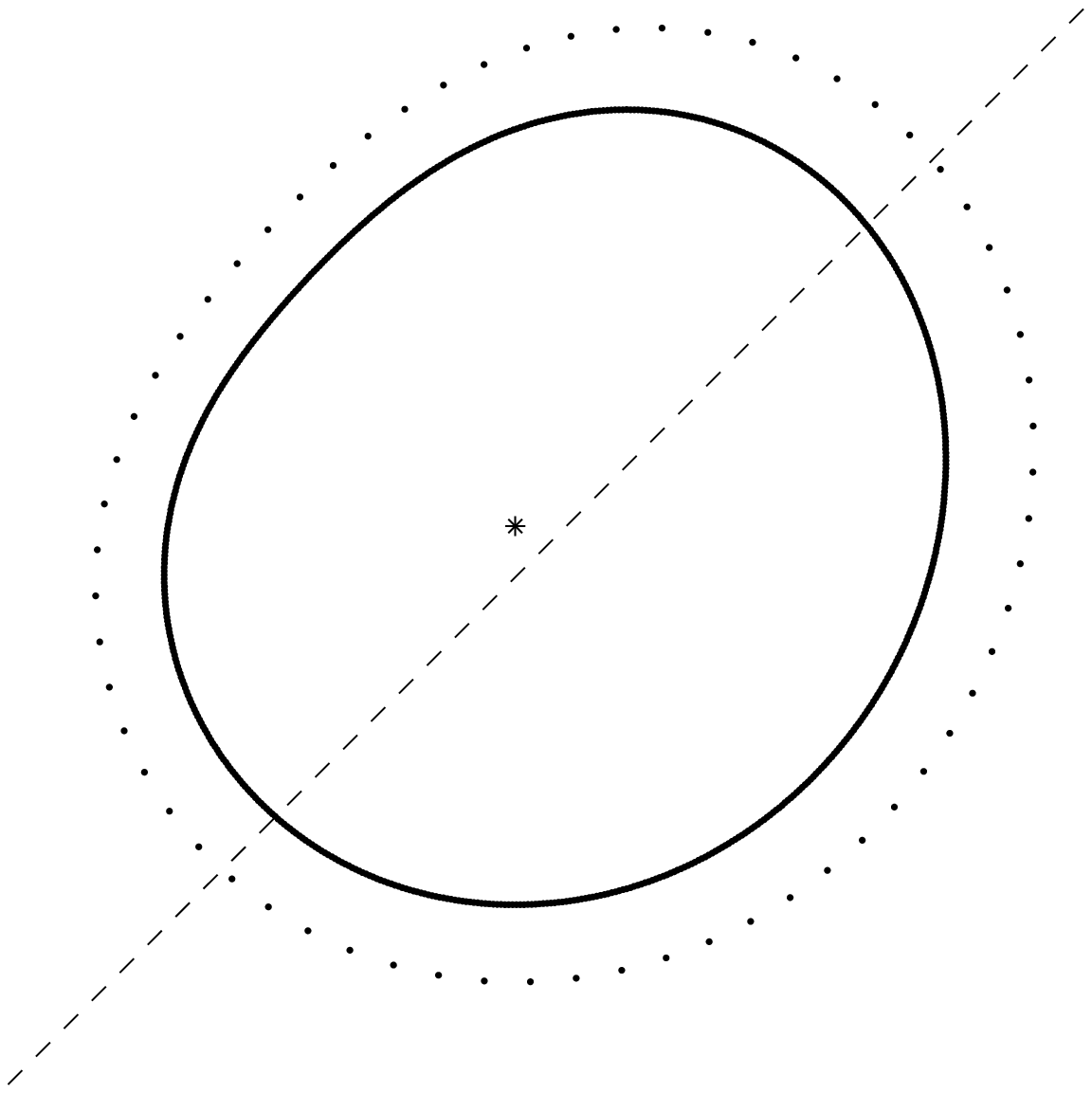,width=5cm}} 
\vspace{.1in} 
{\footnotesize {\bf FIG.1:} 
Full line: The 'Pond' shape~\cite{rmrk}. 
Its radius is roughly $1$, and its perimeter is $L=6.364$. 
Dashed line: The cross section line that is used e.g. in Fig.3c. 
Dots: The 'outer boundary' which is used for the 
BIM tension calculation (see Sec.III-A). 
The transverse distance between 
the actual boundary and the outer boundary 
was chosen in most cases to be $\Delta L \sim 10\Delta s$. 
Star: The selected point which is used 
in Heller's implementation of the PWDM method. } 
\\ \ \\ 

\centerline{\epsfig{figure=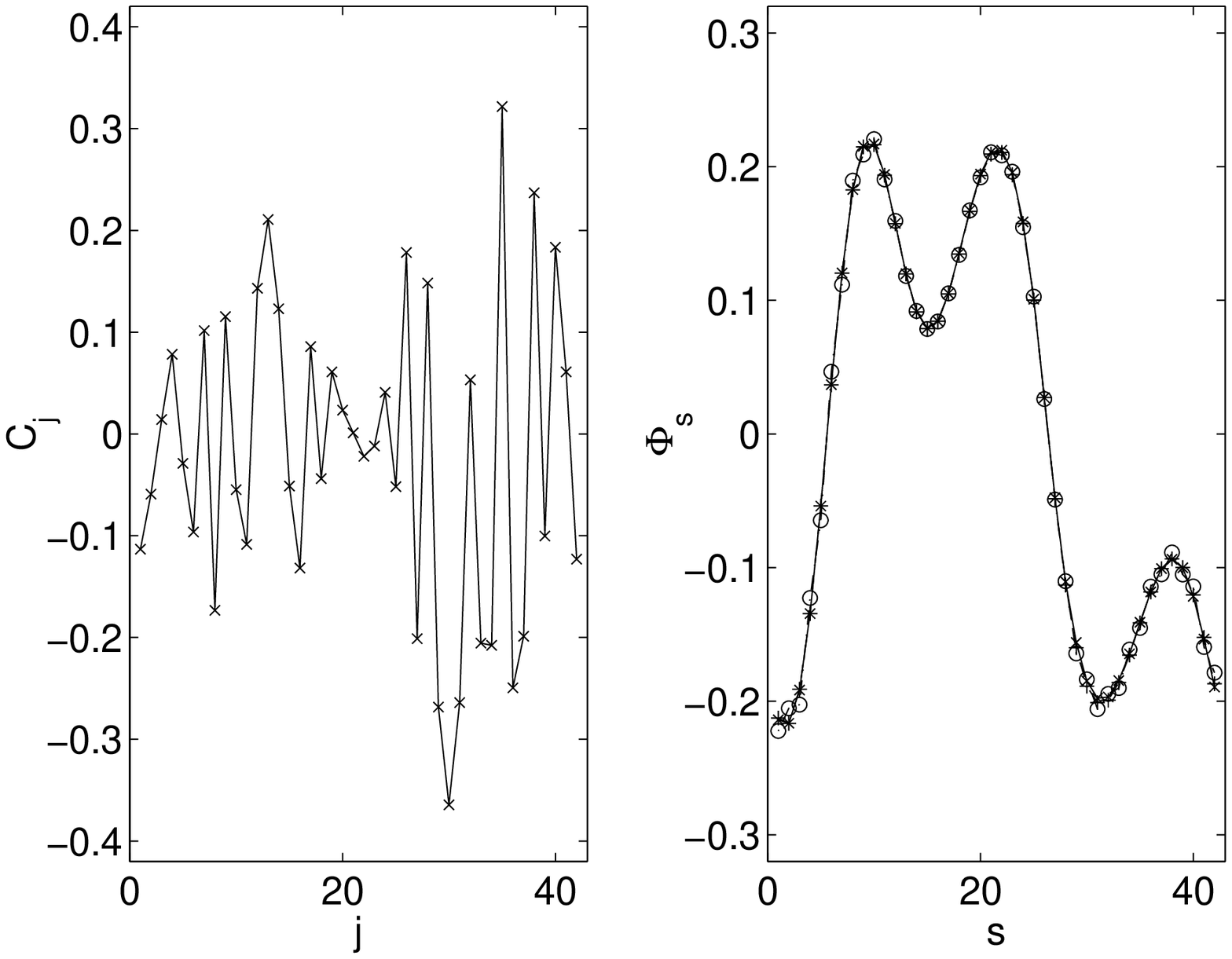,width=0.8\hsize}} 
\vspace{.1in} 
{\footnotesize {\bf FIG.2:} 
The eigenvectors of the Fredholm matrix (Eq.(\ref{e5})) 
are found for $k = k_n = 6.82754592867694$. 
{\bf Right plot:} The right-eigenvector $\Phi$. 
The symbols (x) and (+) and (o) correspond respectively 
to the PWDM choice of Eq.(\ref{e2}), 
to the H1-BIM choice of Eq.(\ref{ea3}), 
and to the J0-GFM choice of Eq.(\ref{e4}). 
{\bf Left plot:} The left-eigenvector ${\mathsf C}$ 
for the PWDM Fredholm matrix.} 

\newpage 
\mpg{2.6cm}{(a) \\ \vspace*{5cm} \\ (b) \\ \vspace*{3cm} }
{\epsfig{figure=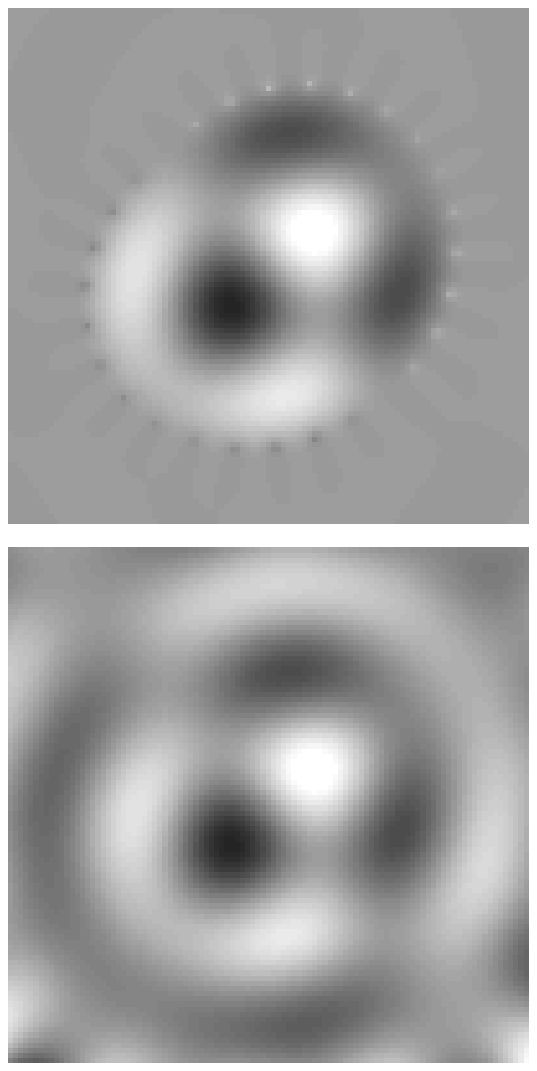,width=0.5\hsize}} 

\mpg{1.4cm}{(c) \\ \vspace*{3cm}}
\epsfig{figure=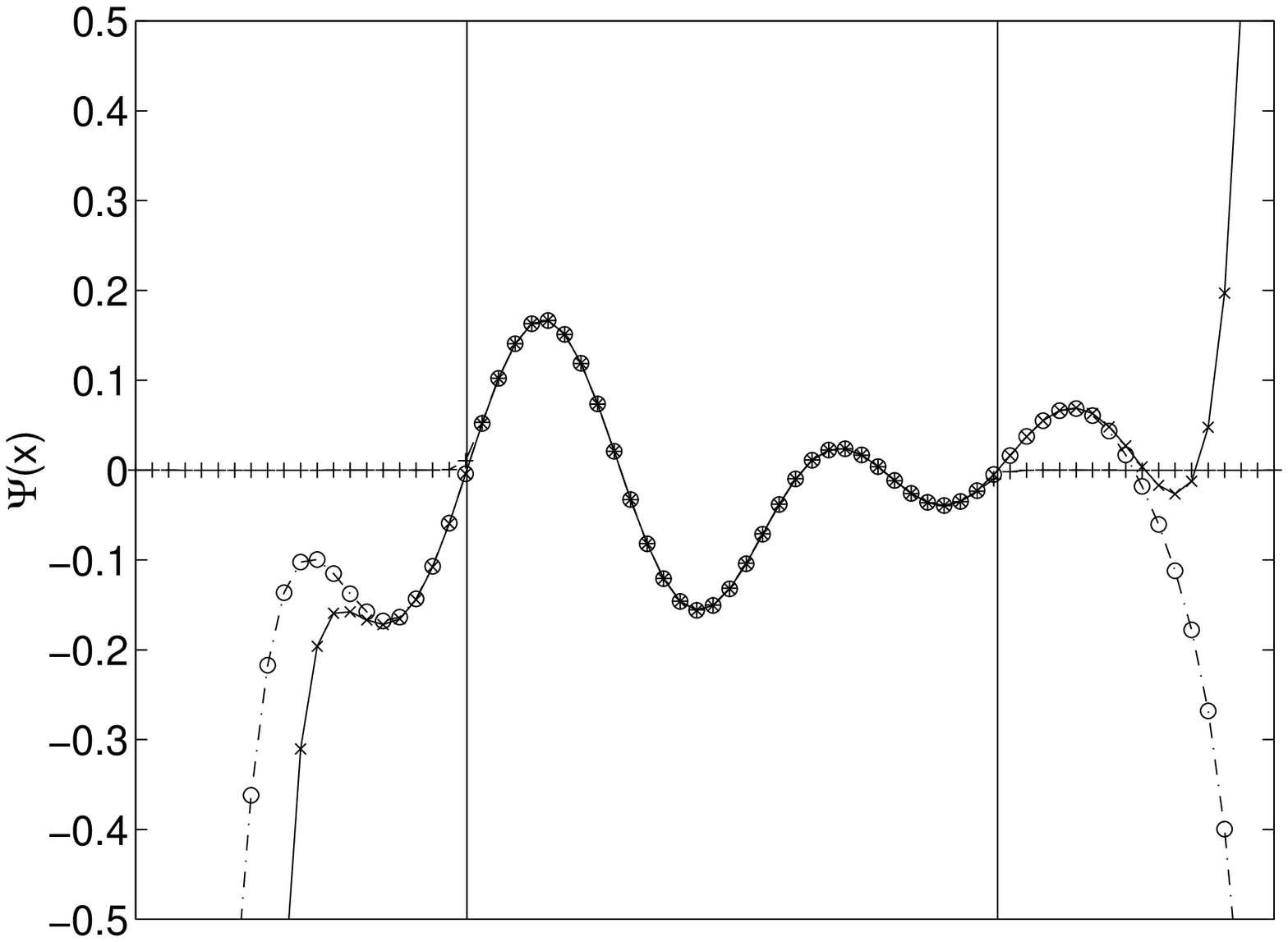,width=0.6\hsize} 

\vspace{.1in} 
{\footnotesize FIG.3abc: See captions in the next page.} 
 
\mpg{0.5cm}{(d) \\ \vspace*{1.7cm}  \\  (e) \vspace*{1.7cm}  \\  (f) \vspace*{1.5cm}}
\epsfig{figure=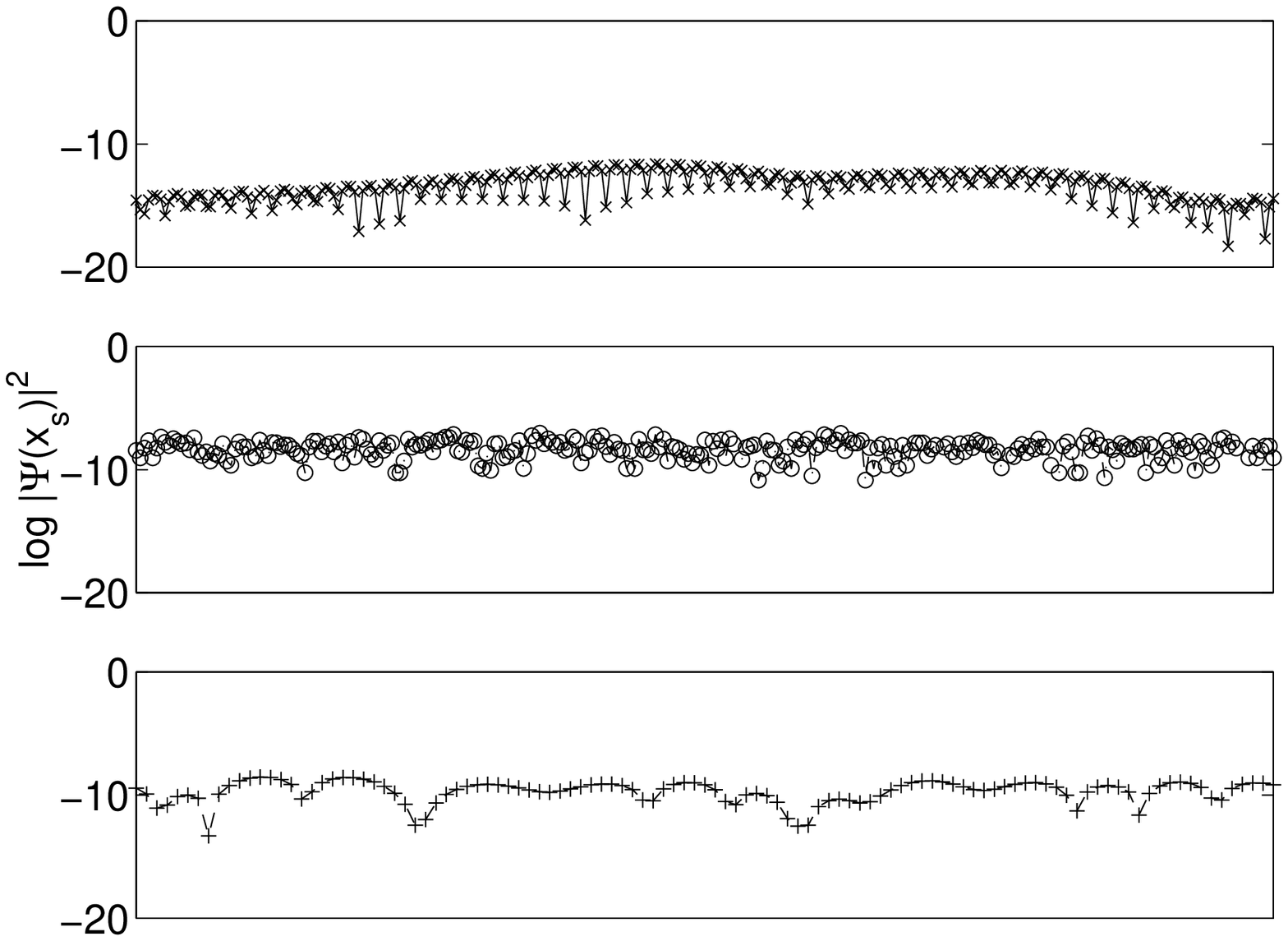,width=0.8\hsize} 
\vspace{.1in} 

{\footnotesize {\bf FIG.3:} 
The eigenfunction at $k=k_n$ is found using 
the eigenvectors of Fig.2. 
(a) An image of $\Psi(x)$ using Eq.(\ref{e7}). 
(b) The same using PWDM and Eq.(\ref{e6}). 
(c) Plot of $\Psi(x)$ along the crossection 
line of Fig.1. The vertical lines indicate 
the location of the boundary. The wavefunctions 
that correspond to images a-b are shown with (+) 
and (x) respectively. We also show with (o) 
the wavefunction that is obtained using J0-GFM 
and Eq.(\ref{e6}). 
Panels d-e-f display plots of $\log(|\Psi(x)|^2)$ 
along the boundary. The symbols are as in c. } 
\\ \ \\ 

\centerline{\epsfig{figure=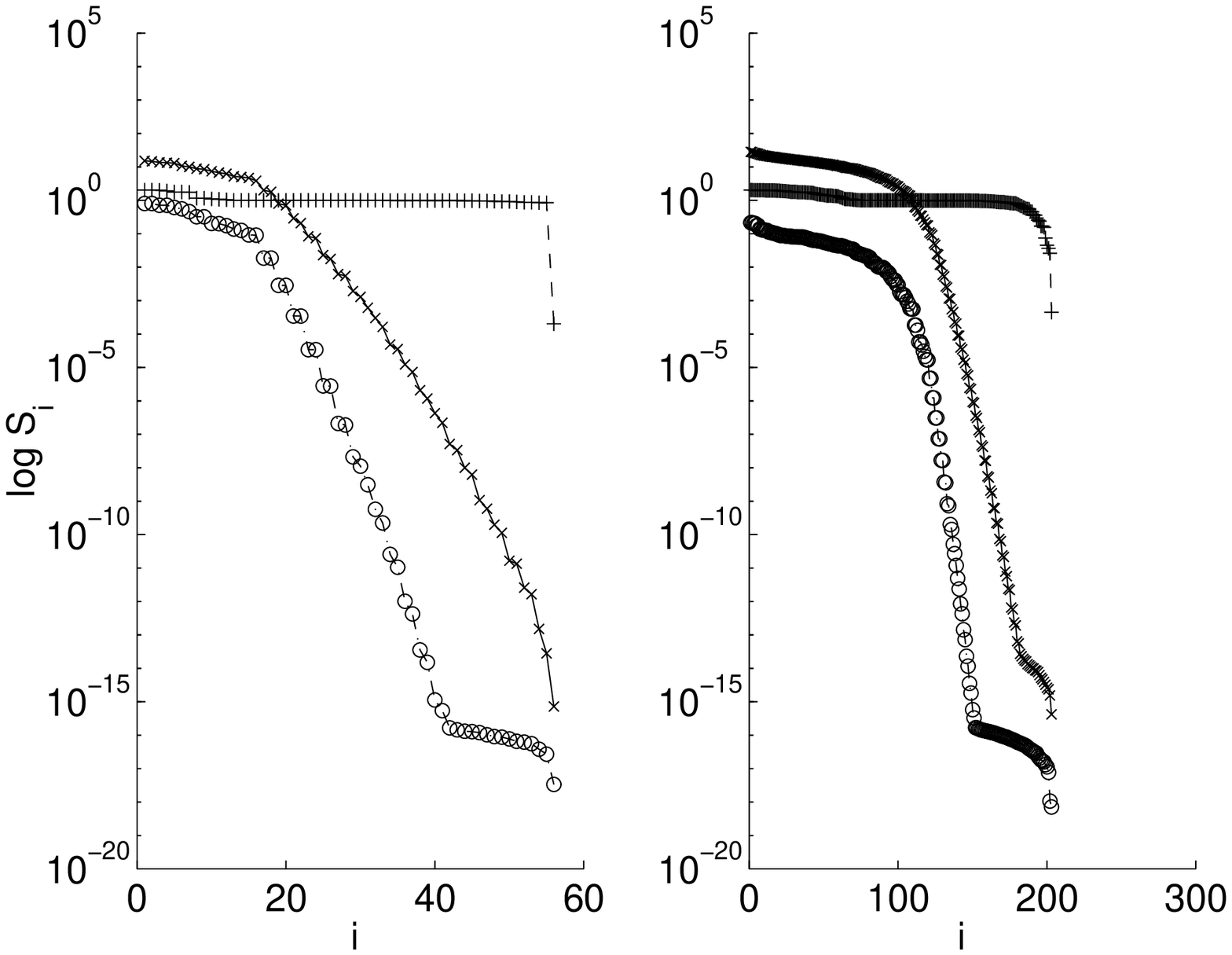,width=0.8\hsize}} 

\vspace{.1in} 

{\footnotesize {\bf FIG.4:} 
Singular values of the Fredholm matrix 
for $k_n=6.82754592867694$ ({\bf left panel}) 
and for $k_n = 50.05474912004408$ ({\bf right panel}). 
The various symbols are as in Fig.2. }  

\newpage
\centerline{\epsfig{figure=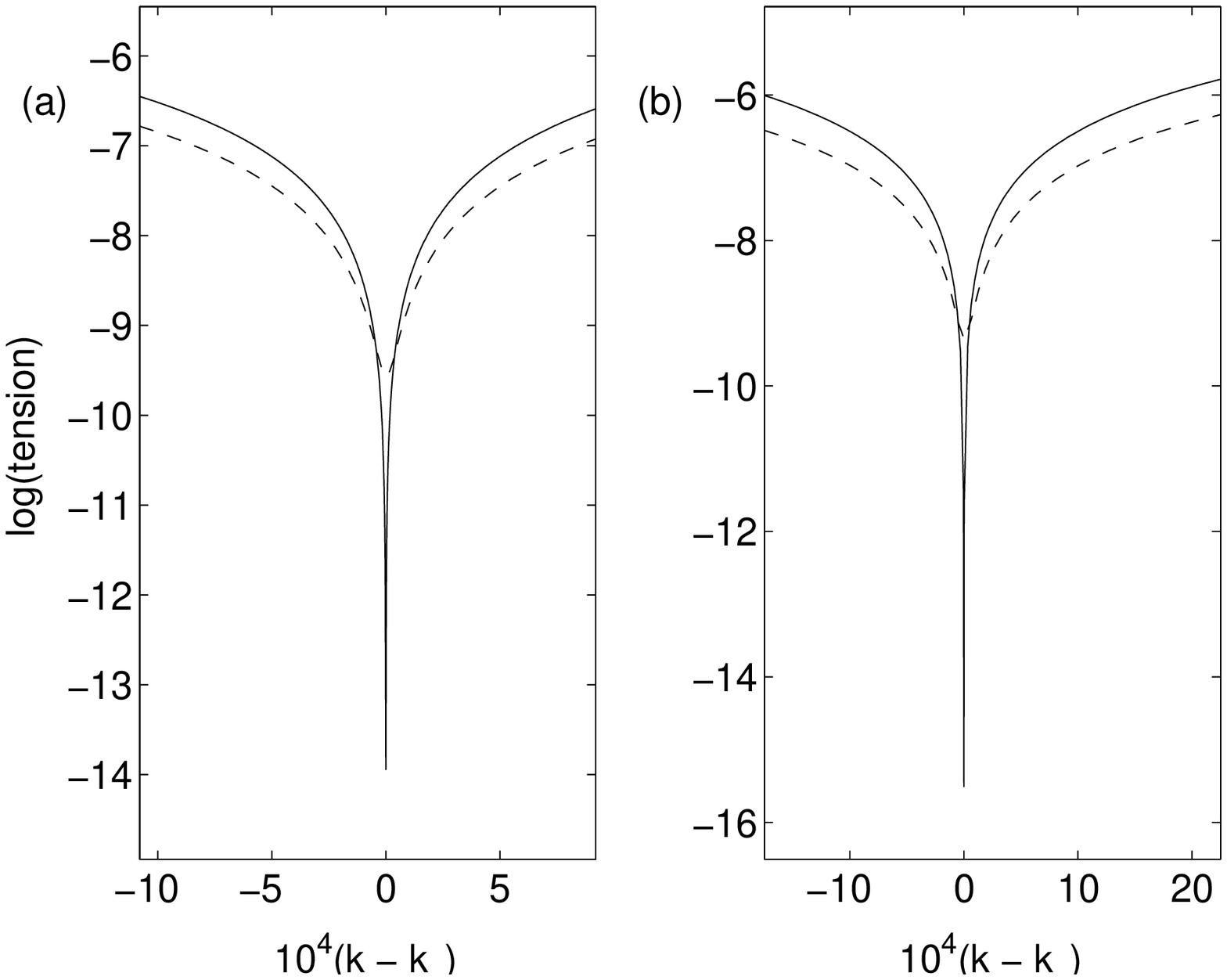,width=0.65\hsize}} 

\mpg{1cm}{(c) \\ \vspace*{4cm}}
{\epsfig{figure=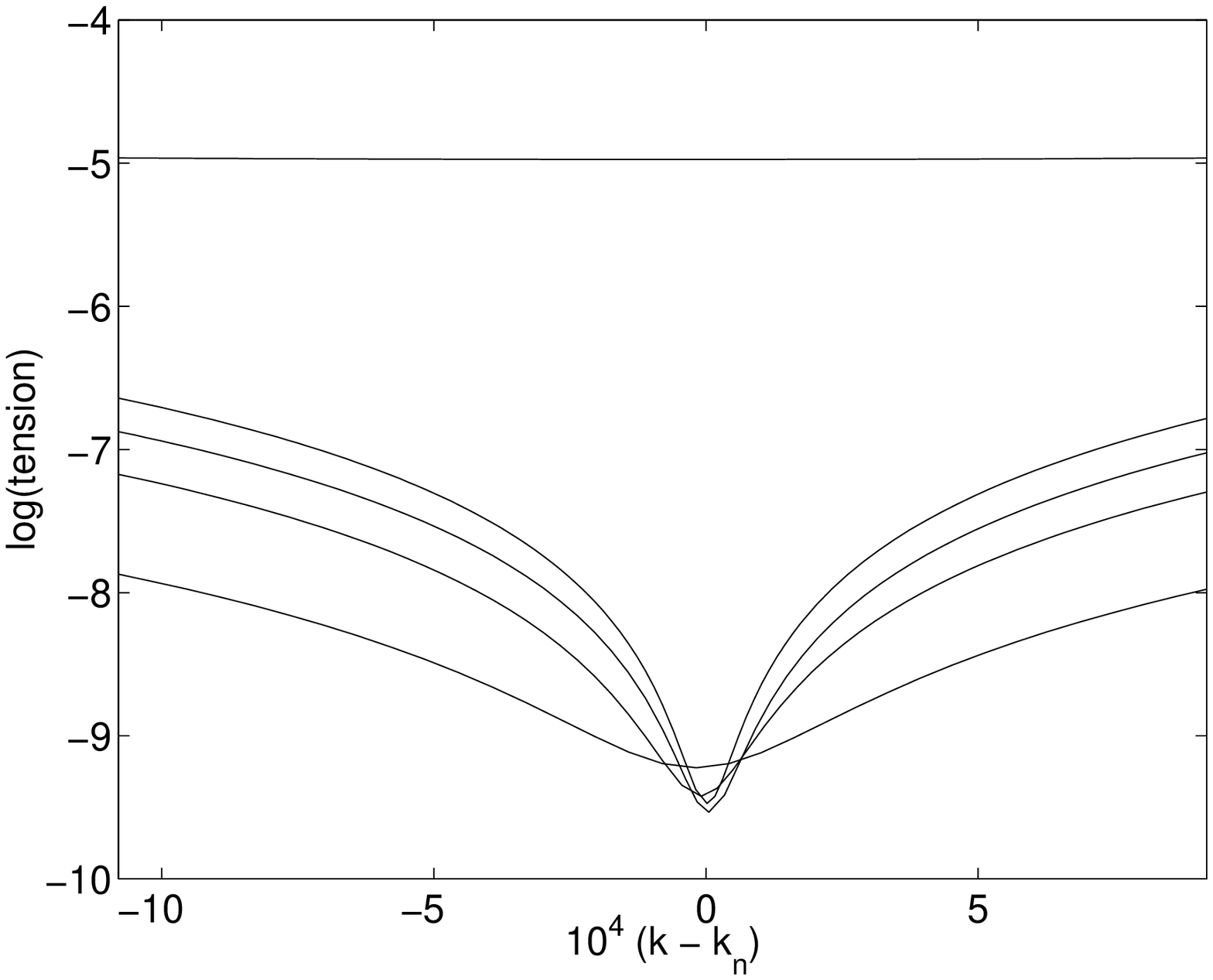,width=0.65\hsize}} 

\mpg{1cm}{(d) \\ \vspace*{2cm}}
{\epsfig{figure=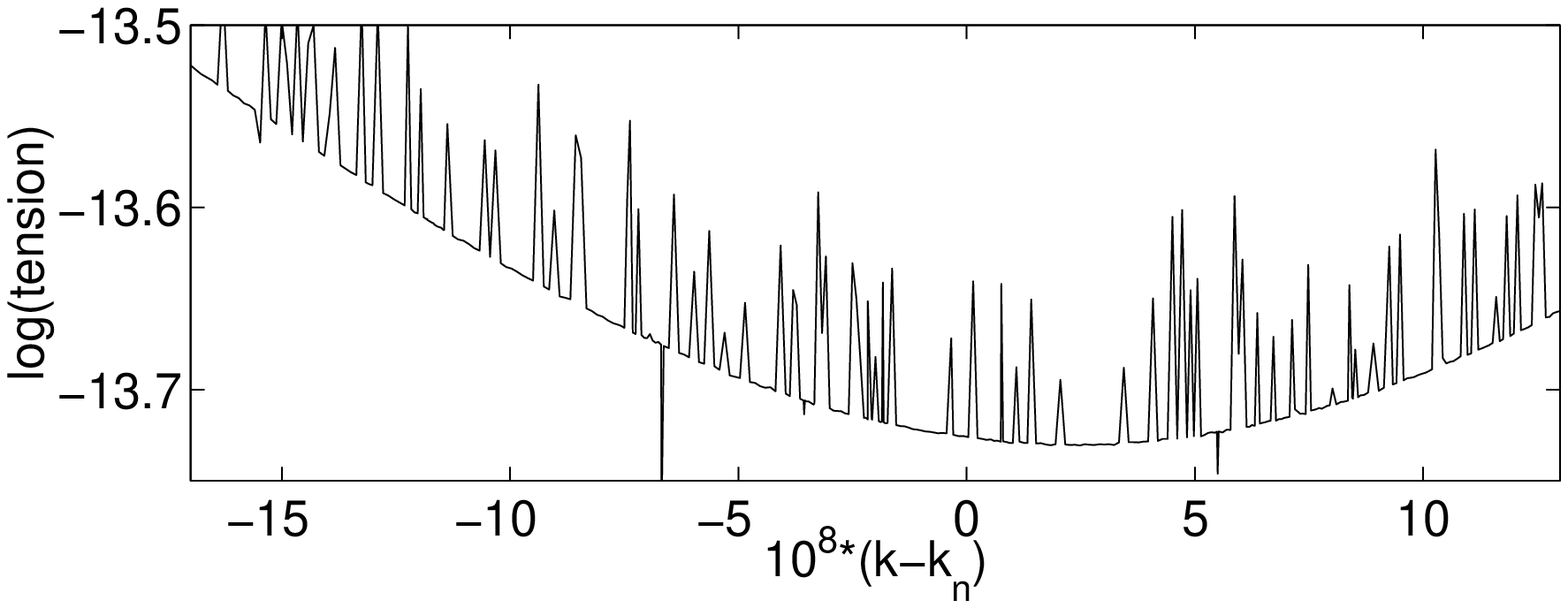,width=0.65\hsize}} 

\vspace{.1in} 

{\footnotesize {\bf FIG.5:} 
The tension as a function of $k$ in the vicinity 
of $k_n=10.14707971517264$ (panels (a),(c),(d)), 
and of $k_n=50.05474912004408$ (panel(b)). 
In the upper panels (a-b) the full line is for the 
PWDM constructed wavefunction, while the dashed line 
is for the BIM constructed wavefunction. 
For the low $k$ state we chose $b=4$, 
while for the high $k$ we used $b=2$. 
Panel (c) demonstrates the dependence 
of the BIM tension on the choice of the distance $\Delta L$. 
The different curves (from the upper to lower) 
correspond to $\Delta L/\Delta s = 1,8,16,12,4$. 
Panel (d) is a zoom over the PWDM minimum. } 

\newpage 
\centerline{
\epsfig{figure=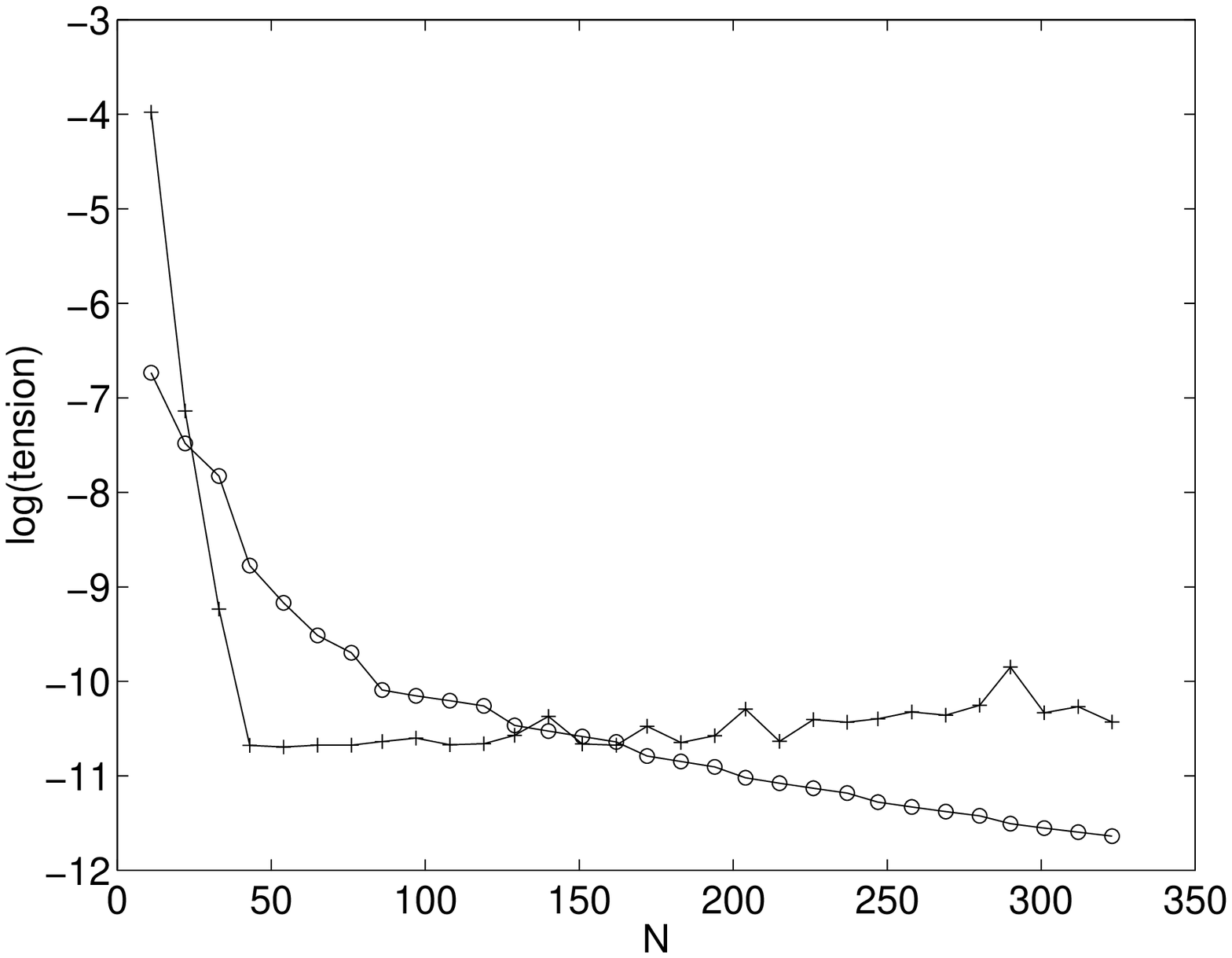,width=0.5\hsize} 
\epsfig{figure=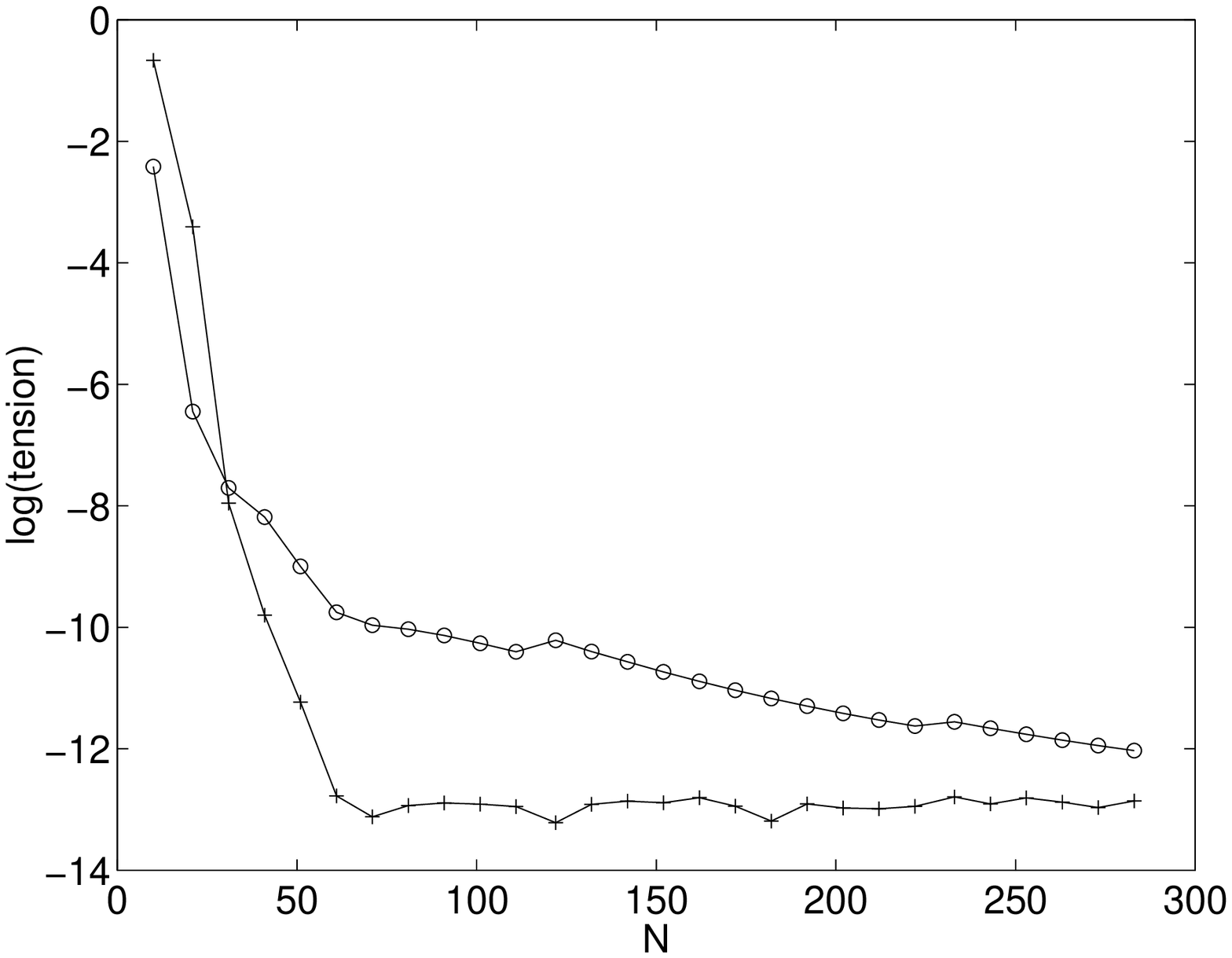,width=0.5\hsize}
} 

\vspace{.1in} 

{\footnotesize {\bf FIG.6:} 
The tension for the constructed eigenstate 
versus the number $N$ of basis functions. 
The {\bf left panel} is for the 
$k_n = 2.40425657792391$ eigenstate, 
and the {\bf right panel} is for the 
$k_n = 6.82754592867694$ eigenstate. 
The symbols (o) and (+) are for 
the BIM and for the PWDM, respectively.} 
\\ \ \\ 

\mpg{0.45\hsize}{ 
\centerline{\epsfig{figure=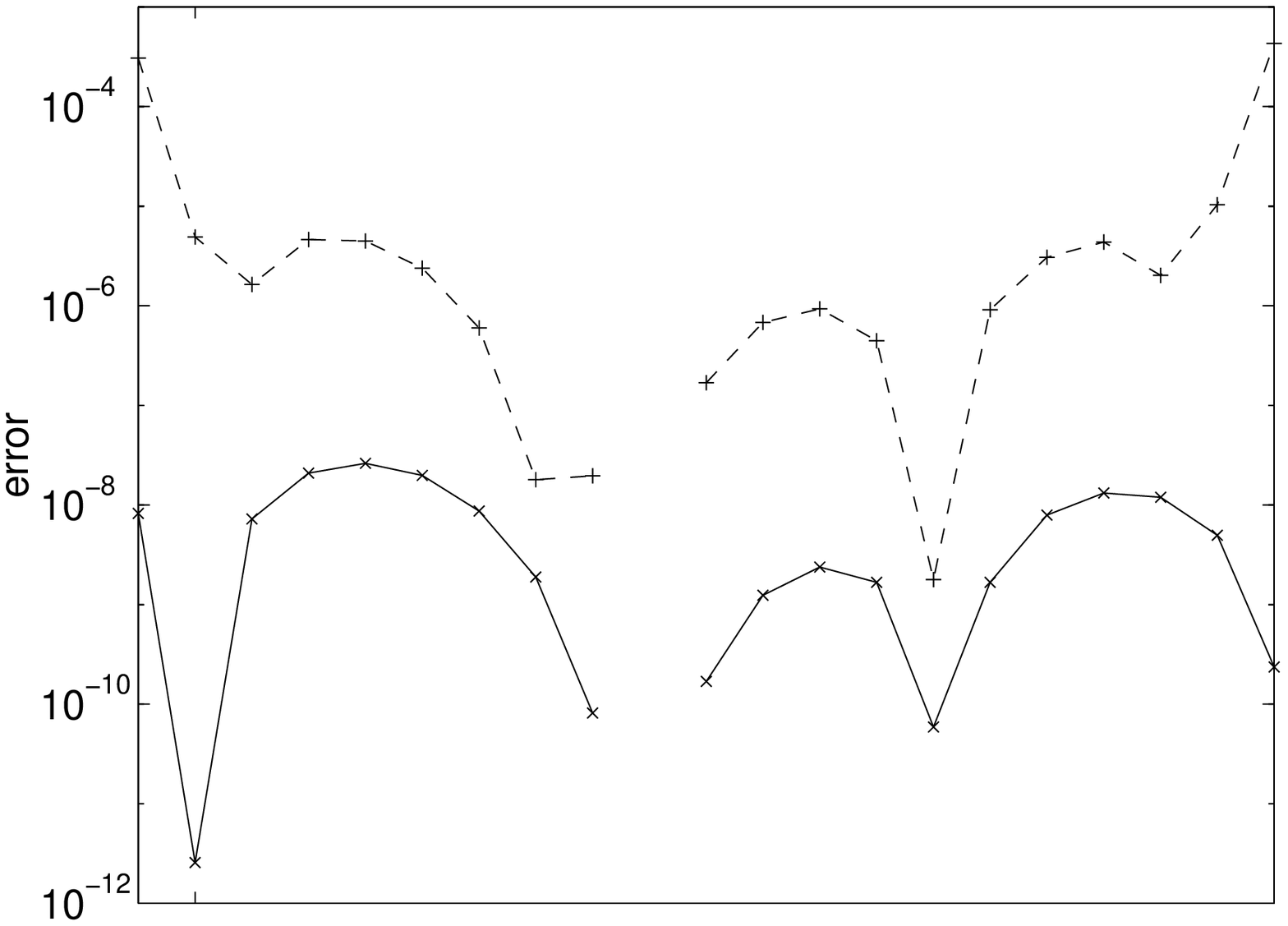,width=\hsize}} 

\vspace{.1in} 

{\footnotesize {\bf FIG.7:} 
Plot of the error $|\Psi_r - \Psi_{exct}(X_r)|^2$, 
along the cross-section line of Fig.1. 
We refer here to the $k_n = 6.82754592867694$ eigenfunction. 
The numerically `exact' wavefunction 
is our best PWDM-constructed wavefunction ($N=69$) 
with tension$=10^{-13}$. 
The `non-exact' wavefunction is either 
BIM-constructed (+) or PWDM-constructed (x), 
with tension $\approx 10^{-8}$. 
In the middle point the error is zero 
by construction (see explanation in the text).} 
\\ \ \\ \ \\ 
} 
\ \ \ \ \ \ \ 
\mpg{0.45\hsize}{ 
\centerline{\epsfig{figure=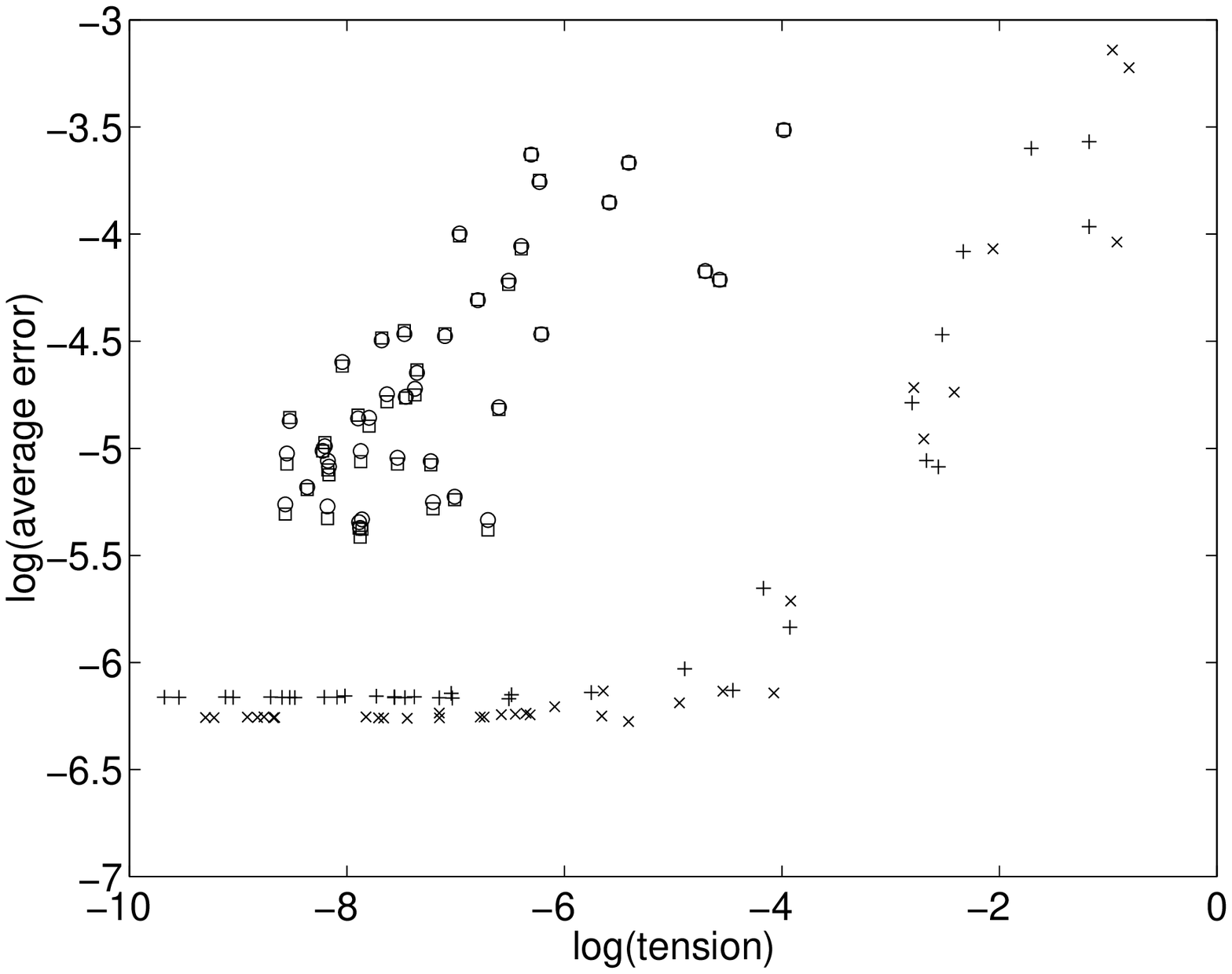,width=\hsize}} 

\vspace{.1in} 

{\footnotesize {\bf FIG.8:} 
The averaged error in the 
determination of the wavefunction, 
versus the tension for the 
$k_n =  6.82754592867694$ eigenfunction. 
For the BIM-constructed wavefunction 
we use squares and (o), 
while for the PWDM one 
we use (+) and (x). 
The error has been determined with respect to
the `exact' wavefunction $\Psi_{exct}$. 
The latter is numerically defined as either 
the best BIM eigenstate (squares and (+)) 
or as the best PWDM one ((o) and (x)). 
For both choices $\Psi_{exct}$ 
had a tension better than $10^{-10}$. } 
\\ \ \\
}

\newpage 

\mpg{0.7cm}{(a) \\ \vspace*{1.7cm}  \\  (b) \vspace*{1.7cm}  \\  (c) \vspace*{1.5cm}}
{\epsfig{figure=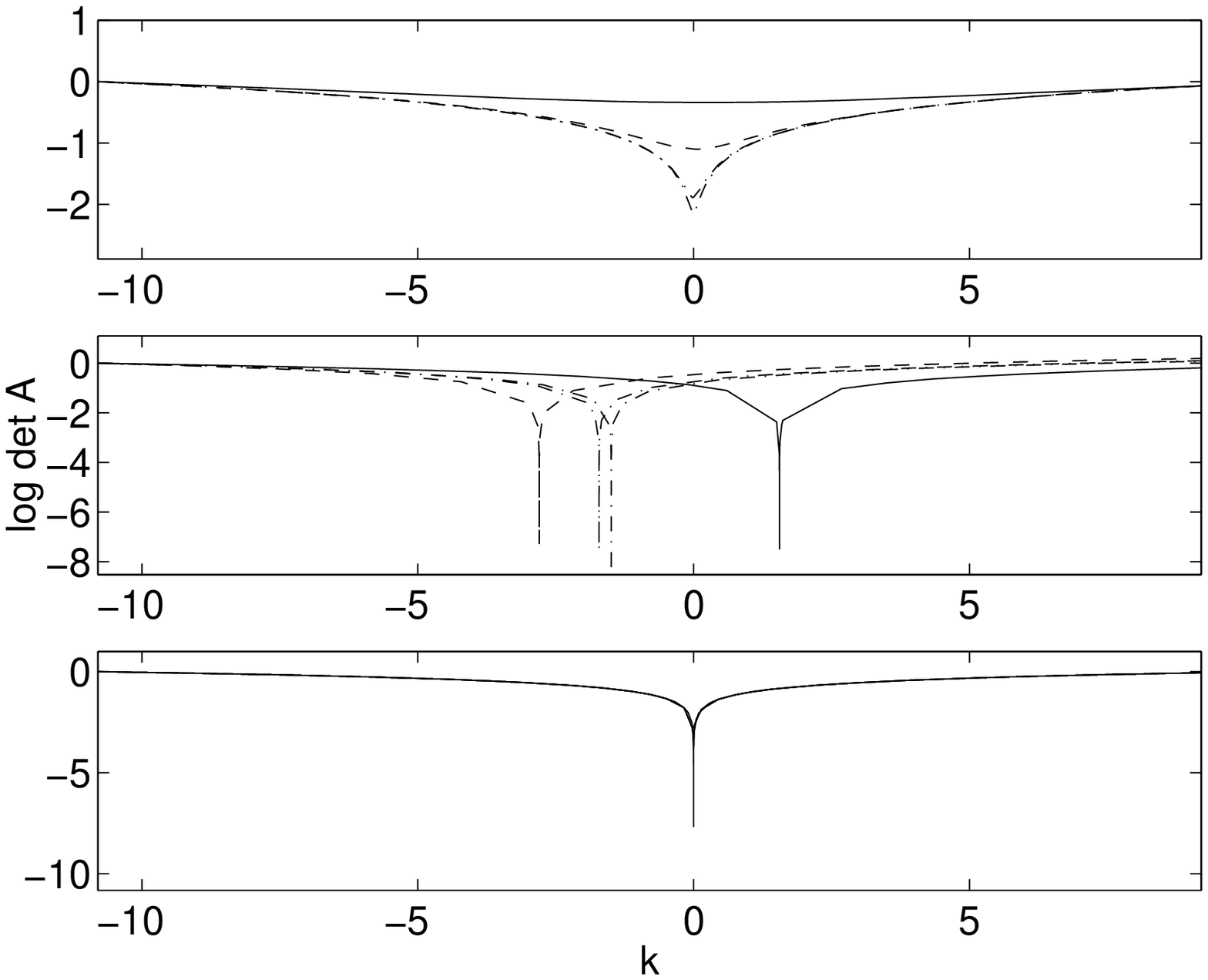,width=0.72\hsize}} 

\mpg{1.5cm}{(d) \\ \vspace*{1.7cm}}
{\epsfig{figure=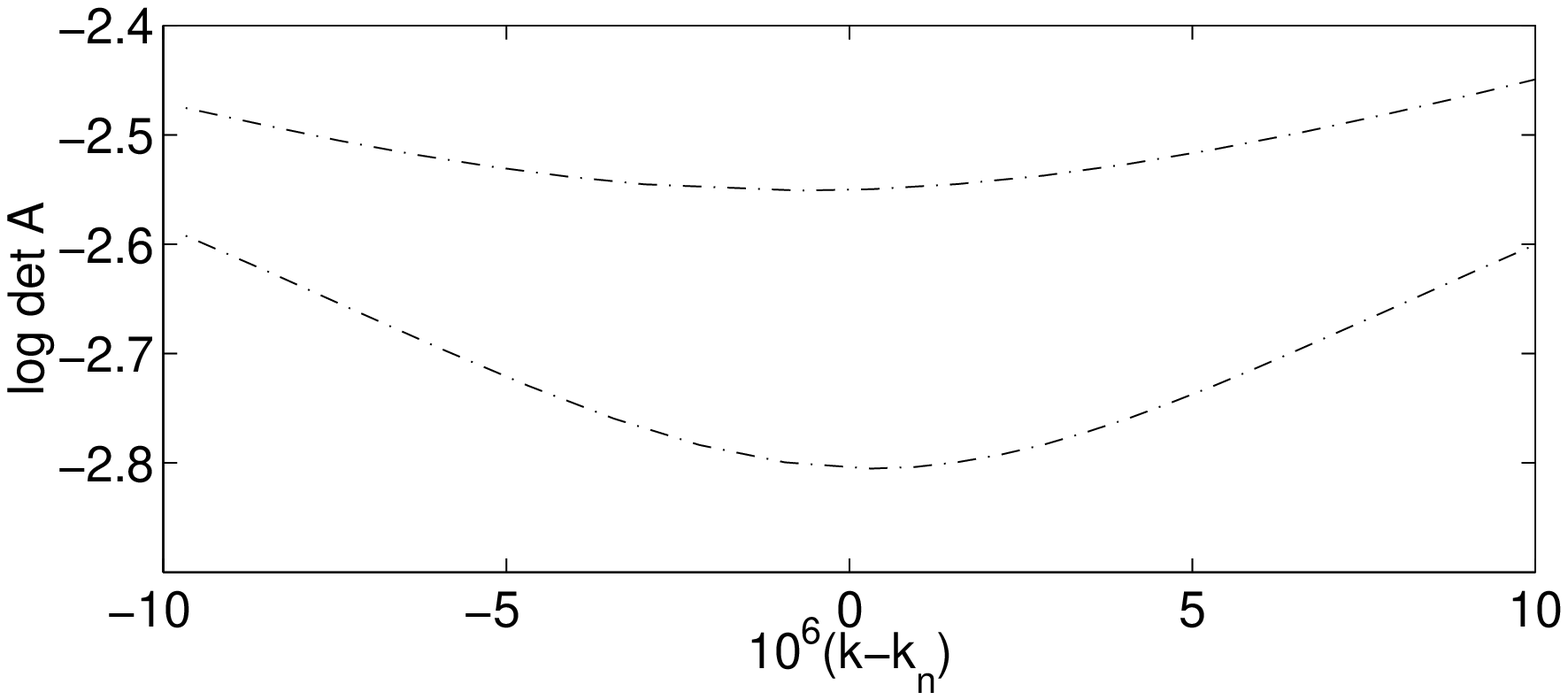,width=0.65\hsize}} 

\mpg{1.5cm}{(e) \\ \vspace*{1.7cm}}
{\epsfig{figure=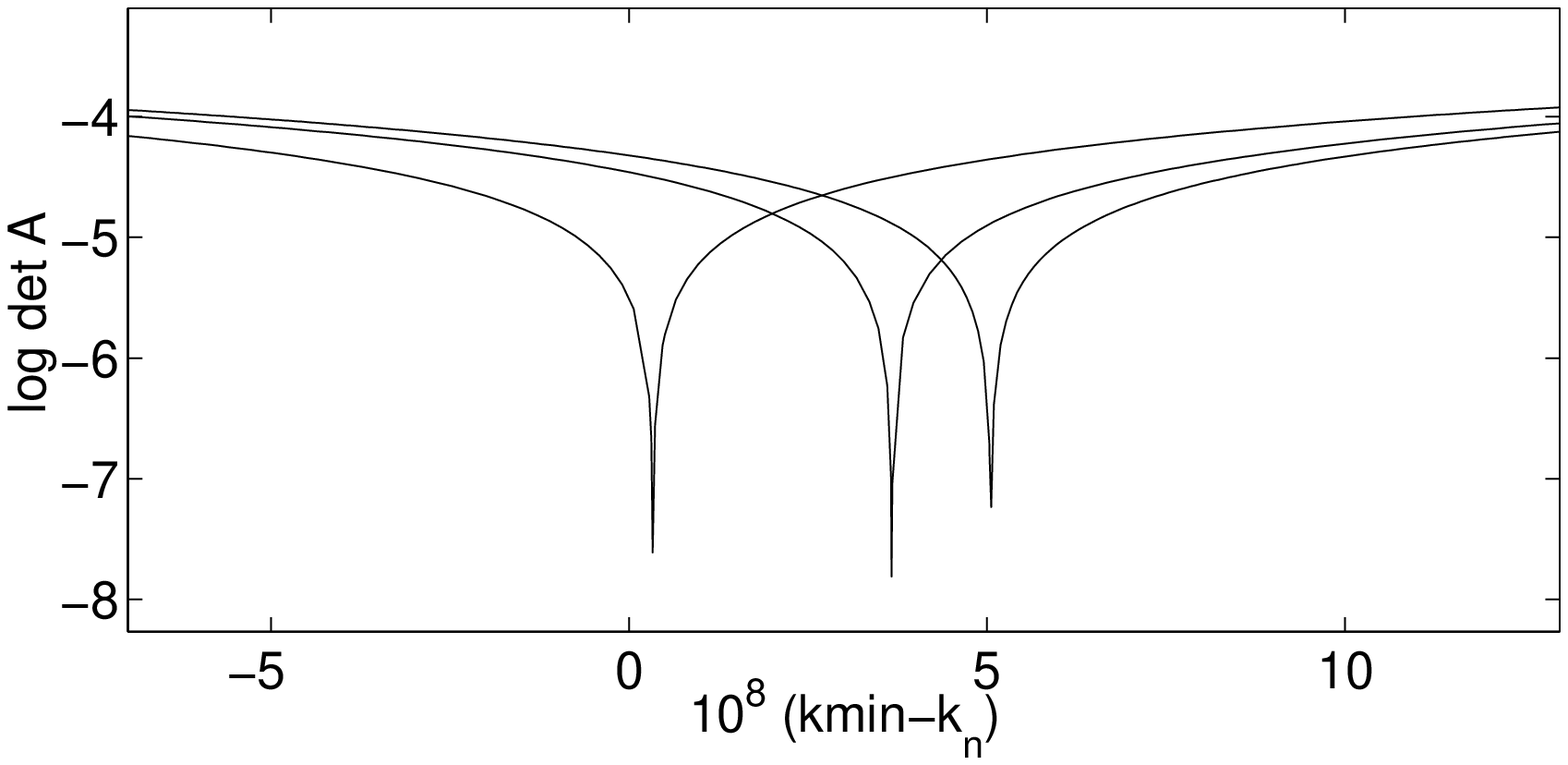,width=0.65\hsize}}

\vspace{.1in} 

{\footnotesize {\bf FIG.9:} 
The Fredholm determinant ($S(k)$) 
versus $k$ in the vicinity 
of $k_n = 10.14707971517264$. 
Note that $S(k)$ is normalized such 
that $S(k)=1$ away from the minima.
Panels a-b-c show $S(k)$
in the cases of the H1-BIM, the Y1-BIM and the PWDM, respectively. 
The lines plotted, 
in order of decreasing $S(k)$ minimum, 
correspond to $b=2,3,4,8$ in panel (a), 
$b=4,8,13,12$ in panel (b) 
and $b=2$ in panel (c). 
The PWDM run in panel 
(c) was repeated 3 times with different values of the 
randomly chosen plane wave phases. 
Panels (d) and (e) give a zoom over 
the minima of panels (a) and (c), respectively. 
We witness some numerical instabilities for both 
the Y1-BIM and the PWDM, though in the latter 
case it is much much weaker, and can be resolved 
only in the zoomed plot (panel~(e)). 
For larger $b$ values, the PWDM instability 
is enhanced, and the results are reduced to 
numerical garbage (not shown). } 


\begin{thebibliography}{99} 

\bibitem{sto} 
H.J. Stockmann, "Quantum Chaos : An Introduction" 
(Cambridge Univ Pr 1999). 
 
\bibitem{datta} 
S. Datta, "Electronic Transport in Mesoscopic Systems" 
(Cambridge Univ Pr 1997). 

\bibitem{kuttler}
J.R.Kuttler and V.G.Sigillito, "Eigenvalues of the
Laplacian", SIAM Review 26 (1984) 163-193.

\bibitem{backer} A. Backer {\em in} 
"The mathematical aspects of quantum maps",
M.Degli Esposti and S.Graffi (Eds), Springer (2003)
[nlin.CD/0204061].  

\bibitem{barnett} 
A.~Barnett, PhD thesis (Harvard, Sept. 2000). 

 
\bibitem{conformal} 
M. Robnik, J. Phys. A {\bf 17}, 1049 (1983). 
 
\bibitem{reichl} 
G.A. Luna-Acosta, Kyungsun Na, and L.E. Reichl, 
Phys. Rev. E {\bf 53}, 3271 (1996). 
 
\bibitem{berry1} 
M.V.~Berry and M. Wilkinson, 
Proc. R. Soc. London A {\bf 392}, 15 (1984). 
R.~E. Kleinman and G.~F. Roach, 
SIAM Review {\bf 16}(2), 214--236 (1974). 
R.~J. Riddell, J. Comp. Phys. {\bf 31}, 21 (1979). 
S.~W. McDonald and A.~N. Kaufman, 
Phys. Rev. A {\bf 37}(8), 3067--3086 (1988). 
 
\bibitem{heller2} 
E.J Heller, {\it Chaos and Quantum Systems}, ed. M.-J.~Giannoni, A.~Voros, 
J.~Zinn-Justin (Elsevier, Amsterdam, 1991), p. 548. 
 
\bibitem{vergini} 
E.~Vergini, PhD thesis (Universidad de Buenos Aires, 1995). 
 
\bibitem{VS} 
E. Vergini and M. Saraceno, Phys. Rev. E, {\bf 52}, 2204 (1995). 
  
\bibitem{li} 
B.~Li, M.~Robnik, B.~Hu, Phys. Rev. E {\bf 57}, 4095 (1998). 
 
\bibitem{bmi} N. Lepore, D. Cohen and E.J. Heller, in preparation. 
 
\bibitem{klaus} 
K. Hornberger and U. Smilansky, 
J. Phys. A {\bf 33}, 2829 (2000). 
 
\bibitem{klaus-long}
K. Hornberger, Spectral Properties of Magnetic Edge States,
Dissertation, Ludwig-Maximilians-Universitat Munchen (2001).
[http://www.mpipks-dresden.mpg.de/~klh/pubs/].

\bibitem{KKR}
M.V. Berry, Annals of Physics {\bf 131}, 163-216 (1981).

\bibitem{boris}
B. Gutkin, archive preprint nlin.CD/0301031.

\bibitem{boasman}
P.A.~Boasman, Nonlinearity {\bf 7}, 485.

\bibitem{dietz}
B.~Dietz and U.~Smilansky, Chaos {\bf 3}, 581 (1993).

\bibitem{uzy}
B. Dietz et al., Phys. Rev. E {\bf 51}, 4222 (1995).

\bibitem{berry2}
M.V.~Berry, J. Phys. A, {\bf 27}, L391 (1994).

\bibitem{lupo}
M.G.E. da Luz, A. Lupu Sax, and E.J. Heller,
Phys. Rev. E {\bf 56}, 2496 (1997).

\bibitem{vega}
J.L. Vega, T. Uzer, J. Ford, Phys. Rev. E {\bf 52}, 1490 (1995).

\bibitem{rmrk} The polar equation of the Pond shape is
\mbox{$r=1+0.2*\cos(\theta+0.9*\cos(\theta))$}.

\bibitem{rmrk2} We assume here the usual free space
boundary conditions used in electrostatics.

\bibitem{rmrk3} Note that Eq.~(\ref{e9}) always gives "a solution"
of the GPL equation. This is true irrespective
of the choice of boundary conditions and Green function.

\end{thebibliography}
\end{document}